\documentclass[letterpaper,twocolumn,10pt]{article}
\usepackage{usenix}

\usepackage{tikz}
\usepackage{amsmath}

\usepackage{filecontents}

\usepackage{bm}
\usepackage{subcaption}
\usepackage{booktabs}
\usepackage{multirow}
\usepackage{graphicx}
\usepackage{pifont}
\usepackage{makecell, multirow,diagbox}
\usepackage[linesnumbered,ruled,vlined]{algorithm2e}
\usepackage{algorithmic}
\usepackage{soul}
\usepackage{amsthm}
\usepackage{amssymb}

\renewcommand{\paragraph}[1]{\par\vspace{2pt}\noindent\textbf{#1}}

\begin{document}

\date{}

\newcommand{\methodname}{\texttt{ATLAS}}

\title{\Large \bf \methodname{}: Automated Approximation of Transformers for Efficient Homomorphic Inference in One Hour}

\author{
{\rm Jianhang Xie$^1$, Sicheng Tan$^2$, Vishnu Naresh Boddeti$^3$, Zhichao Lu$^1$}\\
$^1$City University of Hong Kong \quad 
$^2$Shandong University \quad
$^3$Michigan State University\\
{\rm \texttt{jianhang.xie@my.cityu.edu.hk, sqtsc@mail.sdu.edu.cn,}}\\
{\rm \texttt{vishnu@msu.edu, zhichao.lu@cityu.edu.hk}}\\
} 

\maketitle

\begin{abstract}

Fully homomorphic encryption (FHE) lets a server run inference on encrypted data with strong privacy guarantees, but running a Transformer under FHE is expensive. Its non-linear operations, such as softmax, normalization, and activation, must be replaced with polynomial approximations that the CKKS scheme supports, and the depth of these approximations dominates inference cost. Existing FHE Transformers use hand-tuned approximation settings, such as iteration count and polynomial degree, applied uniformly across layers, models, and tasks. Hand-tuning is slow and error-prone. Even a single uniform setting has about $10^7$ choices, and manual search cannot exploit layer-wise variation. 
AutoFHE, the only automated method with multi-objective search, targets ReLU-only CNNs and needs full fine-tuning per candidate, which is too costly for Transformers. Per-layer settings also push the search space to about $10^{85}$ for BERT and ViT and $10^{228}$ for LLaMA3, beyond both manual and fine-tuning-based search. We present \methodname{}, a training-free framework that automates this search by treating each layer's approximation setting as a multi-objective optimization over latency and accuracy. The problem is hard: the decision space is large (96 or 256 variables), each configuration takes 70 to 1,000 seconds to evaluate even in cleartext, and 85 to 90 percent of configurations are invalid. \methodname{} handles this with a two-stage optimization strategy and a surrogate model, completing the search in about one hour. Compared to an iterative softmax baseline, \methodname{} cuts multiplicative depth and end-to-end latency by about 35 percent with little accuracy loss, and works across encoder-only, decoder-only, and vision Transformers, complementing parallel work on packing and matrix multiplication. 
\end{abstract}

\section{Introduction}

\begin{figure}[t]
    \centering
    \includegraphics[width=0.48\textwidth]{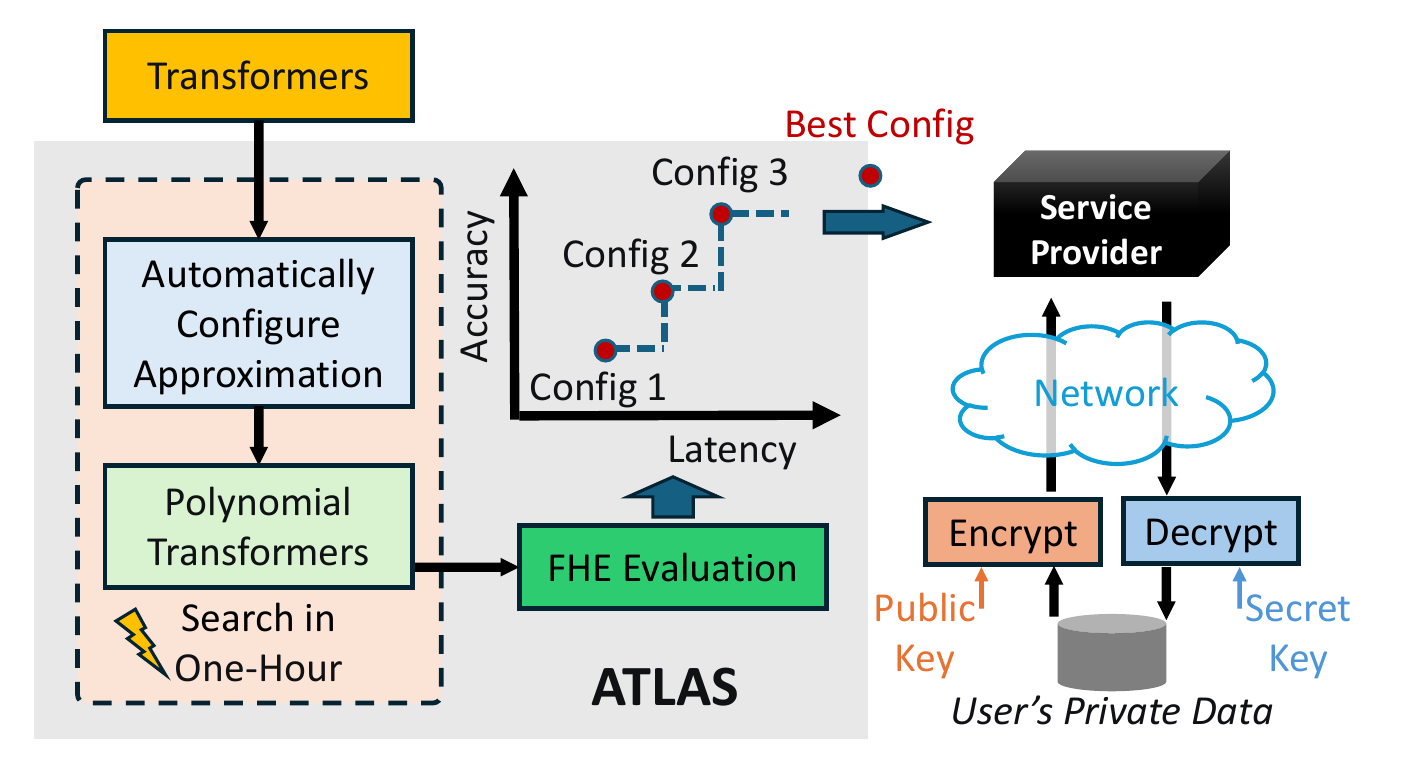} \caption{\methodname{} can automatically design polynomial approximation configurations for FHE compatible Transformers. \methodname{} solutions span the trade-off between accuracy and latency for ciphertext inference and can be deployed on the cloud server to satisfy a range of customer requirements.\label{fig:atlas_intro}}
\end{figure}

The widespread deployment of machine learning has created a growing tension between model utility and data privacy. In sensitive applications such as medical diagnosis, financial assessment, and legal document analysis, a client wishes to obtain predictions from a server-hosted model without revealing the underlying input data. Fully homomorphic encryption offers a compelling cryptographic solution to this problem: as shown on the right side of Figure~\ref{fig:atlas_intro}, under an honest-but-curious server assumption, FHE allows the server to evaluate the model entirely on encrypted data, providing strong privacy guarantees without requiring the server's trust beyond adherence to the protocol.

\begin{figure}[t]
    \centering
    \includegraphics[width=0.48\textwidth]{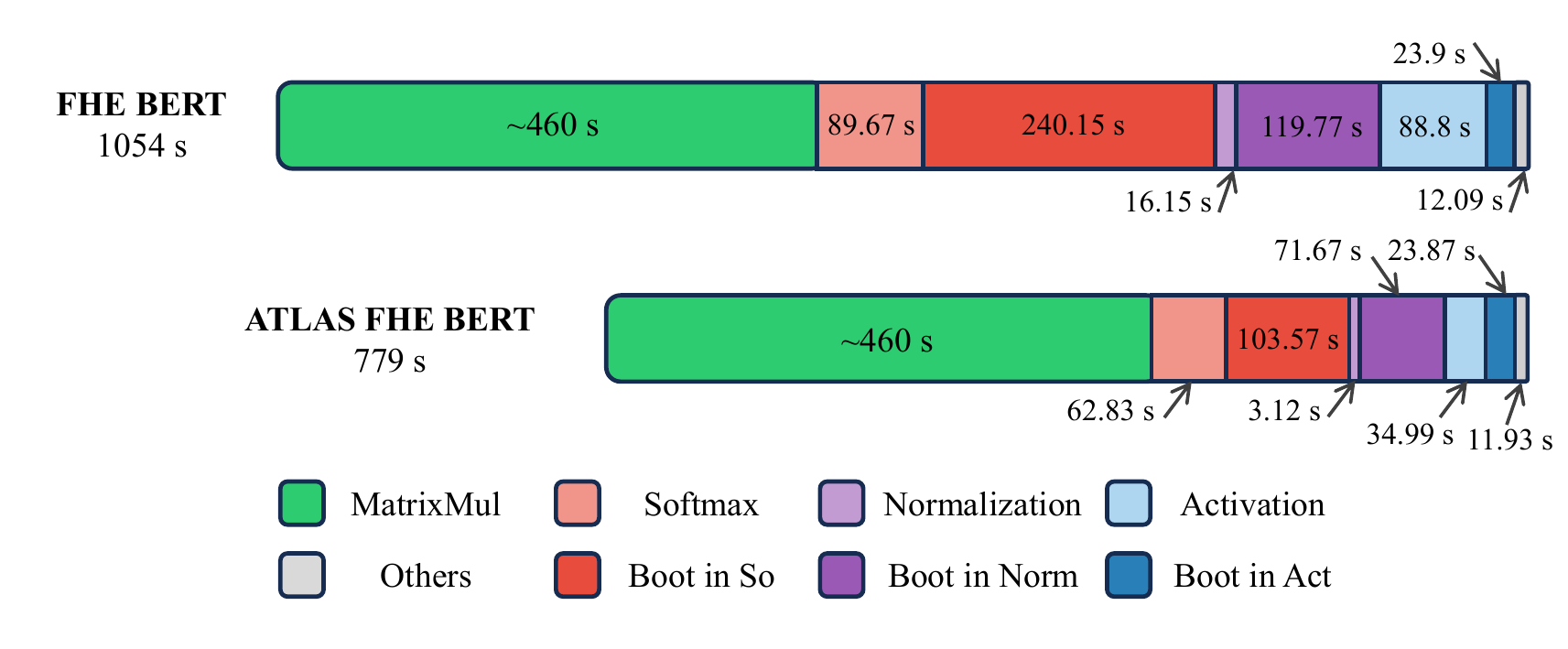} 
    \caption{Runtime breakdown of FHE BERT~\cite{devlin2019bert} on an RTX 4090.
    \methodname{} achieves a ${\sim}{48}$\% saving in non-linear approximation latency, and ${\sim}{26}$\% end-to-end speedup. It saves ${163}$~s, ${61}$~s, and ${54}$~s in softmax, normalization, and activation, respectively, of which ${185}$~s is bootstrapping.
    }\label{fig:time_breakdown}
\end{figure}

Transformer architectures have become the dominant paradigm for such tasks, spanning language understanding (BERT \cite{devlin2019bert}, LLaMA \cite{touvron2023llama}) and vision (ViT~\cite{dosovitskiy2021an}). Deploying Transformers under FHE, however, remains extremely expensive. The Cheon-Kim-Kim-Song (CKKS) scheme~\cite{cheon2017homomorphic,cheon2018full,cheon2018bootstrapping}, which supports approximate arithmetic on real-valued data, is the most practical choice for neural network inference, but it imposes a fundamental constraint: operations are limited to polynomial arithmetic, and the number of sequential multiplications (the multiplicative depth) determines both the ciphertext parameter size and the overall latency. As shown in Figure~\ref{fig:time_breakdown}, a baseline FHE implementation of a Transformer devotes approximately 55\% of inference time to non-linear operations and the bootstrapping they force. This overhead arises because non-linear functions such as softmax, normalization, and activation are incompatible with FHE and must be replaced with polynomial approximations, each consuming significant depth.

Recent work has sought to reduce this cost along three main directions: optimizing linear components such as ciphertext-ciphertext and plaintext-ciphertext matrix multiplication (CCMM \& PCMM)~\cite{guzelhan2026ellmo, lim2025tricycle, moon2025thor}; designing approximations for non-linear operations like softmax, LayerNorm, and GELU~\cite{zhang2024secure,cho2024fast,park2026efficient, badawi2026sok} directly; and improving cryptographic primitives including bootstrapping~\cite{cheon2023homomorphic, bae2024bootstrapping, cheon2025ship, choe2025leveraging, cheon2025grafting} and polynomial multiplication~\cite{kim2024exploring}.

Despite this progress, a critical design decision remains unaddressed across all existing frameworks: how to set the hyperparameters governing each polynomial approximation. 
Current practice treats this as a manual, per-function problem: a practitioner selects an approximation for each non-linear function type (e.g., a degree-$d$ minimax polynomial for GELU) and applies it uniformly across all layers. This is suboptimal because the true objective is to preserve end-to-end task accuracy under FHE constraints, not to approximate individual functions accurately in isolation. Concretely, a uniform setting must satisfy the most demanding layer in the network, so every other layer pays for precision it does not use.

Recent work has begun to automate parts of this problem, but under settings that do not extend to modern Transformers.
AutoFHE~\cite{ao2024autofhe} searches per-layer ReLU degrees in CNNs, but fine-tunes the network for every candidate---a cost that is prohibitive for billion-parameter models.
OLA~\cite{lee26ola} selects layer-wise degrees by dynamic programming, again for CNNs and under a single objective.
SLOTHE~\cite{man26slothe} uses code decomposition to reduce the approximation error of individual non-linearities such as GELU, but leaves softmax and normalization unaddressed and optimizes each function in isolation rather than allocating depth across them.
The core challenge for Transformers is different: configuring every non-linear module per layer---softmax with its iteration count, normalization and activation with their polynomial degrees---under the single global objective of balancing end-to-end accuracy against multiplicative depth.
Optimizing for that objective admits heterogeneous configurations in which different layers use different hyperparameters, trading local approximation fidelity for global depth reduction, but the resulting space is vast: with multiple tunable hyperparameters per operator and dozens of layers per model, valid configurations span $10^{85}$ to $10^{228}$ possibilities depending on the architecture.
It is also expensive to explore---scoring a single candidate takes $77.5$~s for BERT and $1007$~s for LLaMA3-8B even in cleartext---and its feasible region is sparse, as $85$--$90\%$ of configurations drive an approximation outside its valid domain and return NaN.

We present \methodname{}, a framework-agnostic post-processing method that automatically discovers depth-efficient approximation configurations for any FHE-compatible Transformer. \methodname{} formulates the problem as multi-objective optimization over multiplicative depth and end-to-end accuracy, and solves it through a two-stage evolutionary search.
By automating the above process, \methodname{} lowers the barrier to deploying Transformers under FHE by eliminating the need to integrate expertise from cryptography, neural architecture design, and manual hyperparameter tuning. 
A fully automated one-hour search produces a deployable configuration from any pretrained model with no re-training, hand-crafted approximations, or changing packing strategy and encryption parameters.
This makes private inference more accessible to a broader community, from cryptography specialists to artificial intelligence practitioners and secure application developers.
An overview of \methodname{} is shown in Figure~\ref{fig:atlas_intro}. The contributions of this paper are:
\begin{enumerate}
\item \textbf{Problem formulation.} We identify the selection of per-layer polynomial approximation hyperparameters as a critical and underexplored bottleneck in FHE Transformer inference, and formulate it as a multi-objective combinatorial optimization problem over multiplicative depth and end-to-end accuracy.

\item \textbf{\methodname{}.} We present a training-free and framework-agnostic post-processing method that solves this problem through a two-stage evolutionary search, designed to handle conflicting objectives, expensive evaluations, and a high density of invalid configurations. 
\methodname{} is complementary to existing FHE backends (e.g., \texttt{Phantom-FHE}~\cite{yang2024phantom} and \texttt{Desilo-FHE}~\cite{desilo}) and can be applied on top of any FHE-compatible Transformer without modifying the underlying encryption or packing strategy.

\item \textbf{Empirical evaluation.} Evaluated on BERT-base, LLaMA3-8B, and ViT-base, \methodname{} reduces multiplicative depth by ${\sim}35\%$ and inference latency by ${\sim}35\%$ relative to expert-designed baselines, with negligible accuracy loss. 
The entire search completes within 30 minutes for BERT and ViT, and under one hour for LLaMA3-8B, confirming that \methodname{} effectively lowers the barrier to deploying cryptographically secure inference.
\end{enumerate}

\section{Preliminaries}
\subsection{FHE Scheme and Bootstrapping}\label{sec:prelim_ckks_boot} 

\noindent\textbf{CKKS Scheme.} The Cheon-Kim-Kim-Song (CKKS)~\cite{cheon2017homomorphic} is a leveled homomorphic encryption scheme.
Compared with previous FHE schemes, e.g., BGV~\cite{brakerski2014leveled}, BFV~\cite{brakerski2012fully}, and TFHE~\cite{chillotti2020tfhe}, which only support integer number encryption, the CKKS can encrypt real and complex number calculations.

Under CKKS, the plaintexts and ciphertexts are elements in a residue cyclotomic polynomial ring $\mathcal{R}_{\mathcal{Q}_\ell}\!=\!\mathbb{Z}_{\mathcal{Q}_\ell}[X]/(X^N\!+\!1)$.
The modulus is 
${\mathcal{Q}_\ell}\!=\!\prod^{\ell}_{i=0}{q_i}$, where $0\leq\ell\leq D$, the $\ell$ is \emph{level}, the $D$ is level budget.
$N$ is the polynomial degree in CKKS, and a ciphertext has $N/2$ slot counts for single instruction multiple data (SIMD) processing.
We denote the encoding-encryption procedure from cleartext to SIMD ciphertext as $\mathrm{Enc}(\cdot)$.
%
Specifically, assuming that homomorphic addition is $\boxplus$ and homomorphic multiplication is $\boxtimes$, the operations can be described in the following with cleartexts $u$ and $v$ as:
\begin{equation}
\begin{aligned}
    \mathrm{Enc}(u) \boxplus \mathrm{Enc}(v)&\approx \mathrm{Enc}(u+v) \\
    \mathrm{Enc}(u) \boxtimes \mathrm{Enc}(v)&\approx \mathrm{Enc}(u\times v)
\end{aligned}
\end{equation}

\noindent\textbf{Bootstrapping.}
Leveled homomorphic encryption schemes like CKKS only support a finite number of homomorphic multiplications, each of which consumes one level due to rescaling.
So, when the level of a ciphertext becomes zero, the decryption would fail, so we need to perform a \emph{bootstrapping}~\cite{cheon2018bootstrapping} operation to reset to a high level if it is too low to do a computation.
The bootstrapping allows us to evaluate circuit of arbitrary depth, as it homomorphically evaluates the decryption circuit and raises the modulus from ${\mathcal{Q}_0}$ to 
${\mathcal{Q}_D}$
by leveraging the isomorphism 
$\mathcal{R}_{{q}_0}{\cong}\mathcal{R}_{{q}_0}{\times}\mathcal{R}_{{q}_1}{\times}\cdots{\times}\mathcal{R}_{{q}_D}$~\cite{bossuat2021efficient}.
A freshly encrypted ciphertext starts with 
$D$
levels, but bootstrapping consumes $K$ levels and reduces it to 
$D-K$
levels.
%
As bootstrapping requires a lot of key switching operations, it becomes the most time-consuming operation in CKKS.

\subsection{Transformers}
Transformer~\cite{vaswani2017attention} models are typically encoder-only (BERT \cite{devlin2019bert}), decoder-only (e.g., GPT~\cite{radford2019language} and LLaMA~\cite{touvron2023llama}), or vision encoder (ViT \cite{dosovitskiy2021an}). 
Despite this distinction, the structure of the Transformer layer is similar in both encoders and decoders: a stack of $L$ layers.
A basic Transformer layer 
$f(\bm{x})$, includes Attention, Multilayer Perceptron (MLP), and normalization, as shown in Eq.~\eqref{eq:transformer}.
\begin{equation}
\begin{aligned}
f(\bm{x})=\mathrm{Norm2}(\mathrm{MLP}(\mathrm{Norm1}(\mathrm{Attention}(\bm{x}))))
\end{aligned},
\label{eq:transformer}
\end{equation}
where $\bm{x}{\in} \mathbb{R}^{n \times d}$ is the input embedding, $n$ is the number of tokens and $d$ is the hidden size.
For attention, the weight matrices are $\bm{W}_Q, \bm{W}_K, \bm{W}_V, \bm{W}_O \in \mathbb{R}^{d \times d}$, 
yielding $\bm{Q} {=} \bm{x} \bm{W}_Q$, $\bm{K} {=} \bm{x} \bm{W}_K$, and $\bm{V} {=} \bm{x} \bm{W}_V$.
For standard multi-head attention (MHA) with $h$ heads, $\bm{Q}$, $\bm{K}$, and $\bm{V}$ are partitioned into $\bm{Q}_h, \bm{K}_h, \bm{V}_h {\in} \mathbb{R}^{n \times d_k}$, where $d_k {=} d / h$ is the head dimension.
The attention computation for a single head is shown in Eq.~\eqref{eq:mha_head}.
\begin{equation}
\begin{aligned}
\mathrm{Attention}(\bm{Q}_h, \bm{K}_h, \bm{V}_h) = \mathrm{Softmax}\!\left(\frac{\bm{Q}_h {\bm{K}_h}^T}{\sqrt{d_k}}\right) \bm{V}_h
\end{aligned}
\label{eq:mha_head}
\end{equation}
Finally, the outputs of all heads are concatenated and projected via $\bm{W}_O$ to produce the final attention result.
For the MLP, let $d_{\text{ff}}$ be the intermediate size. 
The MLP weights consist of two matrices $\bm{W}_{\text{up}} \in \mathbb{R}^{d \times d_{\text{ff}}}$, $\bm{W}_{\text{down}} \in \mathbb{R}^{d_{\text{ff}} \times d}$, with an activation $\sigma(\cdot)$ placed between them. 
Typically $\sigma(x) {=} \mathrm{GELU}(x)$.
For the LLaMA MLP, $\sigma(x) {=} \mathrm{SiLU}(x)$ and it adopts a gated structure.
For normalization, standard Transformers use LayerNorm~\cite{ba2016layer}.
Modern LLMs such as LLaMA employ RMSNorm~\cite{zhang2019root} to reduce computational cost.
The placement of normalization depends on the specific architecture.

\subsection{FHE Transformers}
Since homomorphic matrix multiplication is nearly lossless in precision, the precision loss arises from the approximation of non-linear components. 
Thus, constructing FHE inference requires a primary focus on these non-linearities.
For Transformers, the non-linear components include \emph{softmax} in Attention, \emph{normalization}, and \emph{activation} in the MLP. 
Compared to CNNs~\cite{lee2022low, ao2024autofhe} where only ReLU requires approximation, building a Transformer inference system under FHE is considerably more challenging.
Currently, the approximation of these 
components~\cite{zhang2024secure, rho2025encryption, cho2024fast, moon2025thor, zhang2025moai, yang2025arion, badawi2026sok} relies on expert-designed heuristics and manually configured 
hyperparameters.

\noindent\textbf{Softmax Approximation.}
The standard softmax function is defined as $\mathrm{Softmax}(\bm{x})_i {=} {\mathrm{exp}({x_i})}/{\sum_{j} \mathrm{exp}({x_j})}$, which contains the non-linear exponential $\mathrm{exp}(x)$ and inverse $1/x$.
Some works replace the Attention with HE-friendly alternatives, such as Gaussian-kernel Attention~\cite{rho2025encryption}, activation-based Attention~\cite{zimerman2024converting}, and BPMax~\cite{park-etal-2025-powerformer}, but these replacements all require model retraining.

Recent works~\cite{zhang2024secure, zhang2025moai, yang2025arion} directly approximate $\mathrm{exp}(x)$ and $1/x$ separately over a given interval.
For example, $\mathrm{exp}(x)$ is approximated via limit $\left(1 + x/2^r\right)^{2^r}$~\cite{zhang2024secure, zhang2025moai} for $x \in [-2^r, 0]$, or via a Chebyshev polynomial~\cite{yang2025arion};
and $1/x$ via Goldschmidt iteration~\cite{zhang2024secure, zhang2025moai} or a Chebyshev-initialized Goldschmidt refinement~\cite{yang2025arion}.
To improve numerical stability, the row maximum is subtracted: $\mathrm{Softmax}(\bm{x}) {=} \mathrm{Softmax}(\bm{x} - \max_i x_i)$. 
In FHE, dynamically computing $\max_i x_i$ is prohibitively expensive; thus, a statistical hard-coded constant maximum $c$ is used instead.
Although this separate approximation incurs low multiplicative depth, the effective domains of the $1/x$ and $\mathrm{exp}(x)$ approximations jointly constrain the softmax input range.
From the NEXUS-EndtoEnd~\cite{nexuse2e} codebase, we measure the usable softmax interval to be roughly $[-2.19, 0]$.
As a result, the valid input range may be narrower than the range of $x_i - c$, leading to approximation failure.

A state-of-the-art alternative employs an \emph{iterative softmax}~\cite{cho2024fast} with normalize-and-square strategy.
It assumes that inputs after subtracting the maximum are non-positive, i.e., $\bm{x} {\in} [-M, 0]^n$. The input is scaled by $1/2^k$ and recovered via $k$ iterations. 
Consequently, the effective input range of the iterative softmax can be estimated as $[-2^k \ln n, 0]$.
For its default setting $k{=}5$ and $n{=}256$, this gives a lower bound of roughly $M{\approx} 177$.
So this mitigates the limited-domain problem.
THOR~\cite{moon2025thor} also proposes a similar square-and-normalize approach that reduces the required number of iterations.
The precision of the above methods depends on the approximation degree of the inverse square root (or inverse) polynomial and the number of iterations $k$.
In iterative softmax~\cite{cho2024fast}, these hyperparameters are manually configured based on expert heuristics.

\noindent\textbf{Normalization Approximation.}
LayerNorm requires the \emph{inverse square root} (invsqrt): $\mathrm{LayerNorm}(x_i) = w_i{(x_i{-}\mu)}/{\sqrt{\sigma^2 {+} \epsilon}} + b_i$, where $\mu$ and $\sigma^2$ denote the mean and variance of input $\bm x$, $w_i$ and $b_i$ are affine transform 
coefficients.
RMSNorm is analogous by omitting the mean.
Nearly all existing works~\cite{zhang2024secure, zhang2025moai, yang2025arion} approximate invsqrt via Newton's method~\cite{qu2023improvements}, often combined with Goldschmidt iterations~\cite{goldschmidt1964applications} for high-precision refinement.
For example, the NEXUS~\cite{zhang2024secure} and MOAI~\cite{zhang2025moai} uses Newton steps to compute $1/\sqrt{x}$, followed by Goldschmidt refinement; 
POLARIS~\cite{badawi2026sok} adopts a Chebyshev polynomial to approximate the $1/\sqrt{x}$ directly, and ARION~\cite{yang2025arion} uses Chebyshev polynomial invsqrt with an additional iterative refinement.

The effectiveness of LayerNorm and RMSNorm under FHE hinges on whether the input variance (or root mean square) lies within the valid approximation interval of the invsqrt.
At the same time, the degrees and numbers of iterations in the above methods remain hand-tuned fixed values.

\noindent\textbf{Activation Approximation.}
The GELU function is defined as $\mathrm{GELU}(x) =x\cdot\Phi({x})$, where $\Phi(x)$ is the cumulative distribution function for Gaussian distribution, 
or by its tanh approximation.
Early FHE Transformers~\cite{zhang2024secure, de2025encryptedllm} and some interactive methods~\cite{Pang24bolt, lubumblebee, dong2023puma} use a piecewise GELU approximation, which employs a sign function approximated by minimax composition~\cite{lee2021minimax, lee2023precise} for segments selection.
However, the piecewise GELU approximation is valid only for a small domain, e.g., $x \in [-8, 8]$ in NEXUS~\cite{zhang2024secure}, which is insufficient to cover the input range encountered during Transformer inference.
To address this limitation, recent works~\cite{ebel2025orion, zhang2025moai, yang2025arion, moon2025thor, badawi2026sok} explicitly define an approximate GELU with polynomials.
For example, MOAI~\cite{zhang2025moai} uses a degree-23 polynomial; ARION~\cite{yang2025arion} adopts a degree-255 Chebyshev polynomial; and THOR~\cite{moon2025thor} employs a composite polynomial with degree-31 and degree-27.
For LLaMA's SiLU, MOAI~\cite{zhang2025moai} does not specify a degree, deferring to Orion~\cite{ebel2025orion}.
However, the polynomial degrees are largely determined by heuristics and manual tuning in all these works.

\subsection{Threat Model}
In the context of secure inference, the following threat model is generally assumed~\cite{lee2022low, ao2024autofhe, zhang2024secure}: a customer uploads encrypted data to the Cloud; 
the Cloud service provider then processes the ciphertext using neural networks deployed on its servers and returns the encrypted output to the customer; 
finally, the customer decrypts the result locally using a secret key. 
Under this model, the provider cannot access the sensitive information contained in the customer's input or output data, while the client remains unaware of the details of neural networks.

\section{\methodname{}: Adapting Transformers for FHE}

\methodname{} is an automated, framework-agnostic post-processing method that designs efficient polynomial approximation configurations for FHE Transformers. Rather than applying a single set of hand-tuned hyperparameters uniformly across all layers, \methodname{} assigns each layer its own approximation hyperparameters. We adopt a multi-objective evolutionary search algorithm to navigate the combinatorial space of layerwise approximation hyperparameters and discover configurations that are (i) significantly more efficient than hand-tuned hyperparameters, and (ii) span the accuracy-efficiency trade-off. Specifically, we adopt a two-stage evolutionary procedure that progressively relaxes layer-wise constraints.

\subsection{Problem Formulation}\label{sec:problem_formulate}

Given a cleartext Transformer $f$, we would like to develop its polynomial approximation $\tilde{f}_{\bm{\lambda}}$ by choosing appropriate hyperparameters $\bm{\lambda}=(\bm{\lambda}_{softmax}, \bm{\lambda}_{norm}, \bm{\lambda}_{act})$ to configure softmax $\bm{\lambda}_{softmax}$, normalization $\bm{\lambda}_{norm}$, and activation function $\bm{\lambda}_{act}$ approximations. 
The goal is to minimize latency overhead without sacrificing security, while ensuring that $\tilde{f}_{\bm{\lambda}}$ exhibits similar predictive performance under FHE. 
This can be mathematically formulated as: 
\begin{equation}
\begin{aligned}
\mbox{minimize}_{\bm{\lambda}}~~Lat(\tilde{f}_{\bm{\lambda}}, \mathcal{D}), \\
\mbox{subject to}~~\left|Acc(f, \mathcal{D}) - Acc(\tilde{f}_{\bm{\lambda}}, \mathcal{D})\right|\leq \epsilon_{acc}, 
\end{aligned}
\label{eq:orig_def}
\end{equation}
where $Lat(\cdot)$ measures runtime latency of the approximated model $\tilde{f}_{\bm{\lambda}}$ under FHE, $Acc(\cdot)$ computes its accuracy over a representative set of data samples $\mathcal{D}$, and $\epsilon_{acc}$ signifies the error tolerance which is typically a small number approaching zero. 
Directly solving Eq.~\eqref{eq:orig_def} imposes two main challenges: (1) measuring latency reliably and consistently on hardware is not trivial; 
(2) $Acc(\cdot)$ metric may not always be sensitive to deviations in model outputs due to approximation error. 
Accordingly, a more practical formulation of the problem is as follows. 
\begin{equation}
\begin{aligned}
\mbox{minimize}_{\bm{\lambda}}~~Mul\_{Depth}(\tilde{f}_{\bm{\lambda}}), \\
\mbox{subject to}~~\left|f(x) - \tilde{f}_{\bm{\lambda}}(x)\right|\leq \epsilon~~\forall x \in \mathcal{D}, 
\end{aligned}
\label{eq:orig_prac}
\end{equation}
where $Mul\_{Depth}(\cdot)$ computes the total multiplicative depth and $\epsilon$ controls the tolerance on output deviation of a FHE Transformer model $\tilde{f}_{\bm{\lambda}}$ from its cleartext counterpart $f$. 

Unfortunately, choosing an appropriate $\epsilon$ in Eq.~\eqref{eq:orig_prac} is also not straightforward, as different Transformer models and tasks require different levels of precision (e.g., ViT can tolerate large deviation in outputs without dropping accuracy on ImageNet-1K classification compared to BERT on GLUE benchmark). An over-constrained $\epsilon$ leads to sub-optimal run-time latency, while an under-constrained $\epsilon$ leads to a noticeable drop in model performance.
As opposed to adaptively/carefully tuning $\epsilon$, we propose to formulate the problem as a multi-objective optimization problem
\begin{equation}
\begin{aligned}
\mbox{minimize}_{\bm{\lambda}}~~\left(Mul\_{Depth}(\tilde{f}_{\bm{\lambda}}), MAE(f, \tilde{f}_{\bm{\lambda}})\right),\\
\mbox{where}~~MAE(f, \tilde{f}_{\bm{\lambda}}) = \frac{1}{|\mathcal{D}|}\sum_{x \in \mathcal{D}}\left|f(x) - \tilde{f}_{\bm{\lambda}}(x)\right|, 
\end{aligned}
\label{eq:multi-obj}
\end{equation}
to simultaneously balance run-time latency and approximation precision, circumventing the need of choosing an $\epsilon$, where the MAE is the \emph{mean absolute error} between original and polynomial Transformer encoder/decoder outputs.
Briefly, the Eq.~\eqref{eq:orig_def} is relaxed to Eq.~\eqref{eq:orig_prac} by replacing latency with multiplicative depth and accuracy with MAE; Eq.~\eqref{eq:multi-obj}  then removes the manual threshold $\epsilon$
through a multi‑objective formulation, yielding a Pareto front of multiplicative depth vs. MAE.

\subsection{Search Space and Encoding}\label{sec:encoding} 
A decision vector $\bm{\lambda}$ specifies how softmax, normalization, and activation functions are approximated in each of the $L$ Transformer layer ($L=12$ for BERT-Base and ViT-Base, $L=32$ for LLaMA3-8B).
The decision variables are grouped by operators 
and arranged in ascending layer order (i.e., from Layer $1$ to $L$), as follows, 
\begin{equation}
\boxed{
\begin{aligned}
\bm{\lambda} = \bigl(
&\underbrace{p_{1}^{(1)},\dots,p_{5}^{(1)},\;\cdots,\;p_{1}^{(L)},\dots,p_{5}^{(L)}}_{\displaystyle\bm{\lambda}_{{softmax}}},\\
&\underbrace{p_{{attn}}^{(1)},p_{{mlp}}^{(1)},\;\cdots,\;
            p_{{attn}}^{(L)},p_{{mlp}}^{(L)}}_{\displaystyle\bm{\lambda}_{{norm}}},\\
&\underbrace{p_{{act}}^{(1)},\;\cdots,\;p_{{act}}^{(L)}}_{\displaystyle\bm{\lambda}_{{act}}}
\bigr),
\end{aligned}
}
\label{eq:total_vars}
\end{equation}
resulting in a total of $5L+2L+L=8L$ integer variables.

\vspace{3pt}
\noindent\textbf{Softmax $\bm{\lambda}_{{softmax}}$ Encoding.}
We adopt the iterative method proposed by Cho et al. \cite{cho2024fast} for softmax approximation, with one modification to use Chebyshev polynomials instead of power basis for coefficient estimation. 
Algorithm~\ref{algo:itersoftmax} outlines our procedure.
\begin{algorithm}
\SetAlgoLined
\caption{\textsc{Iter. Softmax}$(x; k, p_1, {\cdots}, p_k)$ in \methodname{}
}\label{algo:itersoftmax}
\KwIn{$x \in [-M,0]^n$, $k \in \mathbb{Z}_{>0}$, $(p_1, \cdots, p_k)$}
\KwOut{$y \approx \mathrm{Softmax}(x)$}
$y \leftarrow \texttt{ChebyPolyEvalExp}(x / 2^k, 15)$\;
\For{$j \gets 1$ \KwTo $k$}{
    $a \leftarrow \texttt{ChebyPolyEvalInvsqrt}(\sum_{i=1}^n y_i^2, 2^{p_j}-1)$ \tcp*{\color{gray}Chebyshev invsqrt with degree $2^{p_j}-1$}
    $y \leftarrow (a \cdot y)^2$\;
}
\Return{$y$}\;
\end{algorithm}

Given an input $\bm{x}\in[-M,0]^n$, the algorithm first divides the input by $2^{k}$ and applies a 15-degree Chebyshev exponential approximation $\texttt{ChebyPolyEvalExp}$.
It then applies $k$ iterations, each consisting of: (i)~summing the squared intermediate values, (ii)~evaluating a $\texttt{ChebyPolyEvalInvsqrt}$ that approximates the invsqrt, and (iii)~multiplying and squaring the result.
The key hyperparameters are the \emph{total number of iterations $k$} and the \emph{degree of the Chebyshev polynomial $p$ in each iteration}.
We embed the choice of $k$ into the degree variables by allowing the $p_j$ to be zero, indicating that the $j$-th and all its subsequent iterations are skipped.

In summary, each layer $i_{\in 1, \cdots, L}$ is assigned five integer variables $(p_1^{(i)}, p_2^{(i)}, p_3^{(i)}, p_4^{(i)}, p_5^{(i)})$, where $p_1^{(i)} \in [1, 7]$ and $p_2^{(i)}, \dots, p_5^{(i)} \in [0, 7]$.
A non-zero value determines the polynomial degree for the invsqrt function approximation in an iteration, and a zero value omits that iteration and all the subsequent ones. 
The upper bound is set to $7$ following the settings from \cite{cho2024fast}, and the lower bound is chosen such that at least one iteration is executed. 
The full softmax configuration contributing $5L$ integer variables is then encoded as:
\begin{equation}
\begin{aligned}
\bm{\lambda}_{softmax} {=} (p_1^{(1)}, \cdots, p_5^{(1)},\; \cdots,\; p_1^{(L)}, \cdots, p_5^{(L)}).
\end{aligned}
\label{eq:lambda_so}
\end{equation}

\noindent\textbf{Normalization $\bm{\lambda}_{{norm}}$ Encoding.}
Both the Attention and MLP modules in a Transformer layer use normalization.
We consider the Chebyshev polynomial approximation of the invsqrt used in normalization~\cite{badawi2026sok, yang2025arion}, but without additional refinement.
The two degrees $p_{attn}$ and $p_{mlp}$ directly control the accuracy--depth trade‑off: larger values improve the approximation quality at the cost of additional multiplicative depth.
We therefore treat \emph{both $p_{attn}$ and $p_{mlp}$ degrees as searchable integer parameters}.
For each layer $i_{\in 1, \cdots, L}$, two integer variables are assigned: attention degree ${p}_{attn} {\in} [1, 9]$, MLP degree ${p}_{mlp} {\in} [1, 9]$, yielding a polynomial degree of $2^{p_{attn}}{-}1$ and $2^{p_{mlp}}{-}1$, respectively.
The normalization variables of layer $i$ and the full configuration vector with $2L$ integer variables are encoded as:
\begin{equation}
\begin{aligned}
\bm{\lambda}_{{norm}} &=
(p_{{attn}}^{(1)}, p_{{mlp}}^{(1)},\;
\cdots,\;
p_{{attn}}^{(L)}, p_{{mlp}}^{(L)}).
\end{aligned}
\label{eq:lambda_ln}
\end{equation}

\noindent\textbf{Activation $\bm{\lambda}_{{act}}$ Encoding.}
The non-linear activation function 
is approximated by a single Chebyshev polynomial~\cite{ebel2025orion, yang2025arion}.
The degree directly determines the multiplicative depth consumed; we therefore include the 
\emph{degree ${p}_{{act}}$ in the search space}
for exploring the accuracy–depth trade‑off.
Each layer $i$ can select an integer ${p}_{act} {\in} [1, 9]$, yielding a polynomial degree of $2^{p_{act}}{-}1$.
The full activation configuration vector is then:
\begin{equation}
\begin{aligned}
\bm{\lambda}_{{act}} = (p_{{act}}^{(1)},\; p_{{act}}^{(2)},\; \cdots,\; p_{{act}}^{(L)}).
\end{aligned}
\label{eq:lambda_act}
\end{equation}

\begin{figure}[ht]
    \centering
    \begin{subfigure}[b]{0.235\textwidth}
        \centering
        \includegraphics[width=0.985\textwidth]{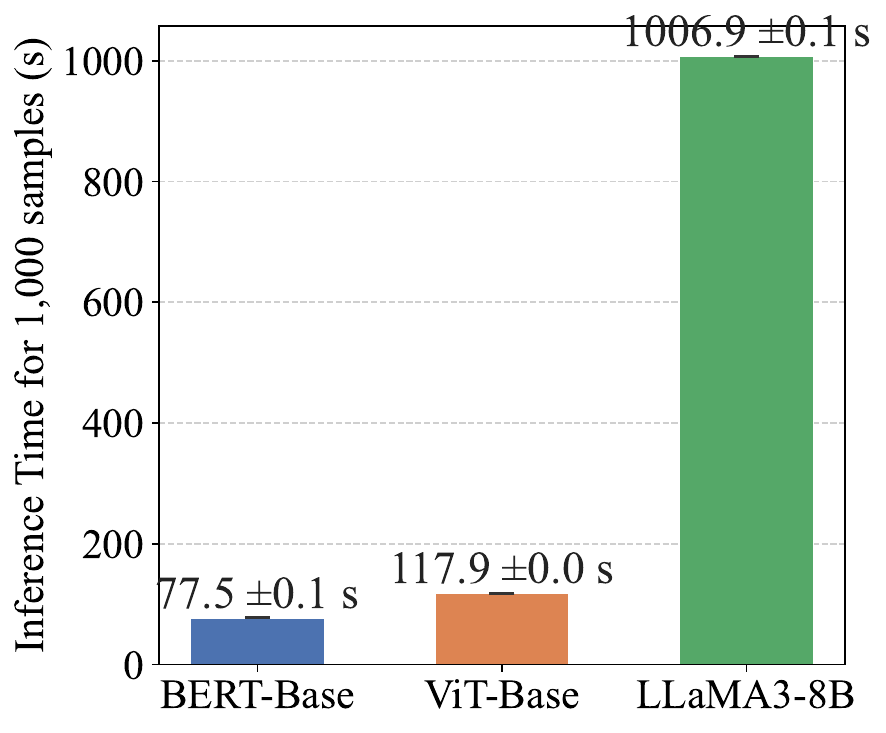}
        \caption{Expensive evaluation}
        \label{fig:challenges_a}
    \end{subfigure}
    \hfill
    \begin{subfigure}[b]{0.235\textwidth}
        \centering
        \includegraphics[width=0.955\textwidth]{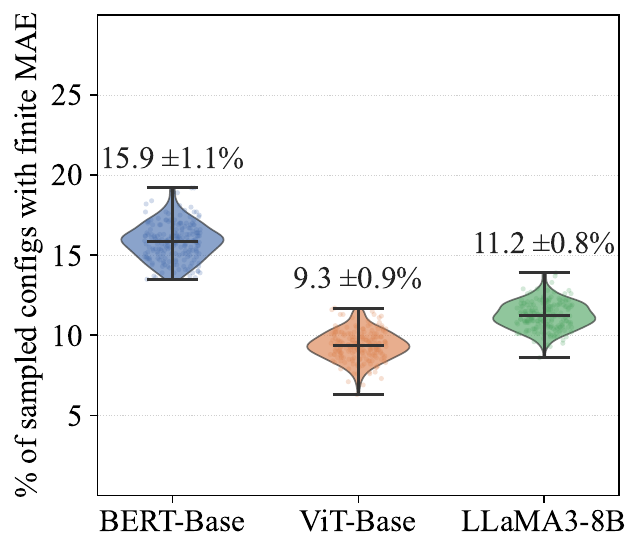}
        \caption{Fraction of finite MAE}
        \label{fig:challenges_b}
    \end{subfigure}
    \caption{Key challenges in searching for FHE approximation configurations. (a) Even a single cleartext evaluation of MAE is time‑consuming, and FHE evaluation is orders of magnitude slower. (b) A large portion of the search space produces invalid (NaN/$\infty$) MAE, resulting in a sparse feasible region.}
    \label{fig:challenges}
\end{figure}

\begin{figure*}[t]
    \centering
    \includegraphics[width=\textwidth]{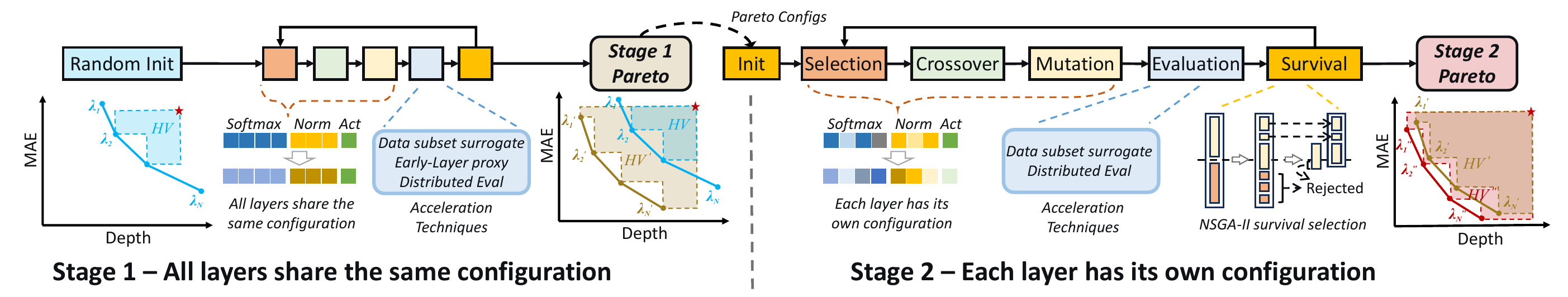} 
    \caption{\methodname{} is a two-stage process. 
    Stage~1 identifies a set of efficient layer‑uniform approximations, and Stage~2 refines these into heterogeneous layer-wise configurations to further improve the depth--accuracy trade‑off.
    }
    \label{fig:atlas_framework}
\end{figure*}

\subsection{Search Challenges \label{sec:challenge}}
The search space defined in Section~\ref{sec:encoding} is combinatorial. 
For a single layer, the number of unique configurations is:
\begin{equation}
\begin{aligned}
|\mathcal{S}_{{layer}}| &= |\mathcal{S}_{{softmax}}| \times |\mathcal{S}_{{norm}}| \times |\mathcal{S}_{{act}}| \\
&= \left(\sum_{k=1}^{5} 7^k\right) \times (9 \cdot 9) \times 9 \\
&= 19\,607 \times 81 \times 9 = 14\,293\,503 \approx 1.43 \times 10^{7},
\label{eq:layer_space}
\end{aligned}
\end{equation}
where $|\mathcal{S}_{{softmax}}| = 19\,607$, $|\mathcal{S}_{{norm}}| = 81$, and $|\mathcal{S}_{{act}}| = 9$.
When configurations are assigned per-layer, the total search volume grows to $|\mathcal{S}_{{layer}}|^{L}$, yielding $14\,293\,503^{12} \approx 7.3 \times 10^{85}$ for $L{=}12$ (BERT/ViT) and $14\,293\,503^{32} \approx 9.2 \times 10^{228}$ for $L{=}32$ (LLaMA3‑8B).
Beyond the sheer size of this space, the optimization must overcome several fundamental difficulties.

\begin{itemize}
    \item \textbf{NP-Hard structure.}
    Selecting per-layer approximation hyperparameters under a depth–accuracy budget is a constrained integer program that generalizes multi-dimensional knapsack and per-layer precision assignment, both NP-hard. 
    No polynomial-time exact algorithm is therefore expected, motivating heuristic, population-based search rather than exhaustive or exact optimization.

    \item \textbf{Conflicting objectives.}
    The two objectives---multiplicative depth and MAE---are inherently conflicting: higher-degree polynomials and more iterations reduce MAE but consume more depth, and vice versa.
    Finding a single solution that simultaneously minimizes both objectives is generally infeasible; instead, we seek a set of Pareto-optimal solutions that represent the efficient trade-off between the two.

    \item \textbf{Expensive evaluation.}
    Each candidate configuration must be instantiated as a model
    and evaluated to obtain its MAE and depth.
    Even in the cleartext polynomial setting used during search, a single function evaluation (FE) is costly for computing MAE.
    As shown in Figure~\ref{fig:challenges}(a), a single FE under cleartext already takes $77.5$~s for BERT, $118$~s for ViT, and $1007$~s for LLaMA; under end-to-end FHE, the latency is even higher.
    This cost compounds in population-based methods, where every generation evaluates an entire population, making naive deployment of evolutionary search prohibitive without aggressive evaluation acceleration.

    \item \textbf{Sparse feasible region.}
    A large fraction of the search space yields configurations that produce NaN or infinite MAE, rendering them useless.
    Figure~\ref{fig:challenges}(b) shows that the proportion of configurations with finite MAE is below $10{\sim}15\%$ for all three models.
    This sparse signal makes the optimization landscape highly discontinuous, further complicating the search.
\end{itemize}

\subsection{Two-Stage Search with NSGA-II}\label{sec:twostage_search}

In \methodname{}, to address the challenges outlined in Section~\ref{sec:challenge}, we develop a two-stage evolutionary optimization strategy based on the Non-dominated Sorting Genetic Algorithm II (NSGA-II)~\cite{deb2002nsga2}.
The core idea is to decompose the search: first explore the compact \textsc{LayerProblem}, where all $L$ layers share the same configuration; then refine them into per-layer settings by solving \textsc{NetworkProblem}, where each layer can have different configurations.  
As shown in Figure~\ref{fig:atlas_framework}, the first stage is computationally efficient and produces a set of layer-uniform yet high-quality configurations that serve as a {warm start} for the second stage, thereby drastically accelerating convergence.

\subsubsection{Preliminaries on NSGA-II}
\methodname{} adopts NSGA-II~\cite{deb2002nsga2} to
handle the
two competing objectives---multiplicative depth and MAE---through
Pareto dominance: $\boldsymbol{\lambda}_1$ dominates
$\boldsymbol{\lambda}_2$ if it is no worse in both objectives and
strictly better in at least one. The non-dominated set forms the
Pareto front, whose quality we measure by the \emph{hypervolume}
(HV)~\cite{zitzler1999multiobjective} with respect to a fixed
reference point; a larger HV in the (multiplicative depth, MAE) space indicates
a front that achieves lower depth, lower error, or a better
trade-off between the two.

NSGA-II evolves a population of solutions (i.e., configuration
vectors) over generations through four operators: \emph{selection},
\emph{crossover}, and \emph{mutation} produce new candidate solutions from the
current population, and \emph{survival selection} then determines
which solutions advance to the next generation. The survival step
is the core of NSGA-II: given a combined pool of current and newly generated
solutions, \emph{non-dominated sorting} partitions the pool into
fronts $\mathcal{F}_1, \mathcal{F}_2, \ldots$, admitting solutions
front by front; when a front cannot be fully admitted,
\emph{crowding distance}---the objective-space distance to
neighbors within the same front---serves as a tiebreaker, favoring
solutions in less densely populated regions. Together, the two
criteria balance convergence toward the Pareto front with diversity
along it, as illustrated in Figure~\ref{fig:atlas_framework}. 
The main evolutionary operators are summarized below.
\begin{itemize}
    \item \texttt{Selection.} Promising solutions, referred to as \emph{parents}, are chosen via binary tournament selection: two individuals are randomly sampled, and the one with better non‑dominated rank is kept. This process repeats until enough parents are selected.
    \item \texttt{Crossover.} A standard two-point crossover is used to exchange sub-components between two parents at each cross-over point to create a new set of solutions, referred to as \emph{offspring}.
    \item \texttt{Mutation.} An integer step mutation operator then perturbs each variable in the offspring solution by +1 or -1 following a binomial distribution with a probability proportional to one over the number of variables, rounding to the nearest integer within the bounds. 

\end{itemize}

\subsubsection{Two-Stage Framework}
However, directly applying NSGA-II to the $8L$‑variable
\textsc{NetworkProblem}
suffers from the curse of dimensionality: the initial population is extremely sparse, and the search wastes many evaluations on infeasible or poor regions.
We observed that a configuration where all layers are identical already yields a reasonable baseline; evaluating it on the
\textsc{LayerProblem}
is fast, and the HV grows quickly.
In contrast, the per‑layer search requires tens of thousands of FEs to reach a comparable HV.
This motivates an efficient two‑stage design:

\begin{itemize}
    \item \textsc{LayerProblem} — all $L$ layers share the same configuration by constraining:
    \begin{equation}
    \begin{cases}
    \bm{p}^{(1)} {=} \bm{p}^{(2)} {=} \cdots = \bm{p}^{(L)}, ~~\mathrm{where}\ \bm{p} {=} (p_1, p_2, p_3, p_4, p_5)\\
    p_{{attn}}^{(1)} {=} p_{{attn}}^{(2)} {=} \cdots {=} p_{{attn}}^{(L)}, ~~p_{{mlp}}^{(1)} {=} p_{{mlp}}^{(2)} {=} \cdots {=} p_{{mlp}}^{(L)} \\
    p_{{act}}^{(1)} {=} p_{{act}}^{(2)} {=} \cdots {=} p_{{act}}^{(L)}.
    \end{cases}
    \label{eq:layer_sharing}
    \end{equation}

    \item \textsc{NetworkProblem} — each layer has its own configuration, i.e.\ the $8L$‑variable bounded integer space represented by $\bm{\lambda}$ in \eqref{eq:total_vars}. 
\end{itemize}

\noindent\textbf{Stage~1.}
NSGA‑II searches the single‑layer configuration that is applied identically to all $L$ layers.
With a moderate budget (e.g., $1/10$ of the total budget), the algorithm produces a Pareto set of high‑quality layer‑uniform configurations $\bm{\lambda}_{layer}$.

\noindent\textbf{Stage~2.}
These $\bm{\lambda}_{layer}$, together with a small subset of randomly created configurations, form a {seeded} initial population of the \textsc{NetworkProblem} for NSGA-II.
Because $\bm{\lambda}_{layer}$ already resides in promising regions of the objective space, the search starts from a favorable basin and efficiently refines per‑layer configurations.
The Stage~2 budget (e.g., $9/10$ of the total budget) is set to match the total evaluation budget.

\subsection{Accelerating the Search}\label{sec:accelerate}
Even with the two-stage strategy, each evaluation call remains expensive.
To further reduce the cost of the search, we introduce two surrogate models that approximate the true objectives at a fraction of the cost and a distributed evaluation technique.

\begin{figure}[t]
    \centering
    \begin{subfigure}[b]{0.235\textwidth}
        \centering
        \includegraphics[width=\textwidth]{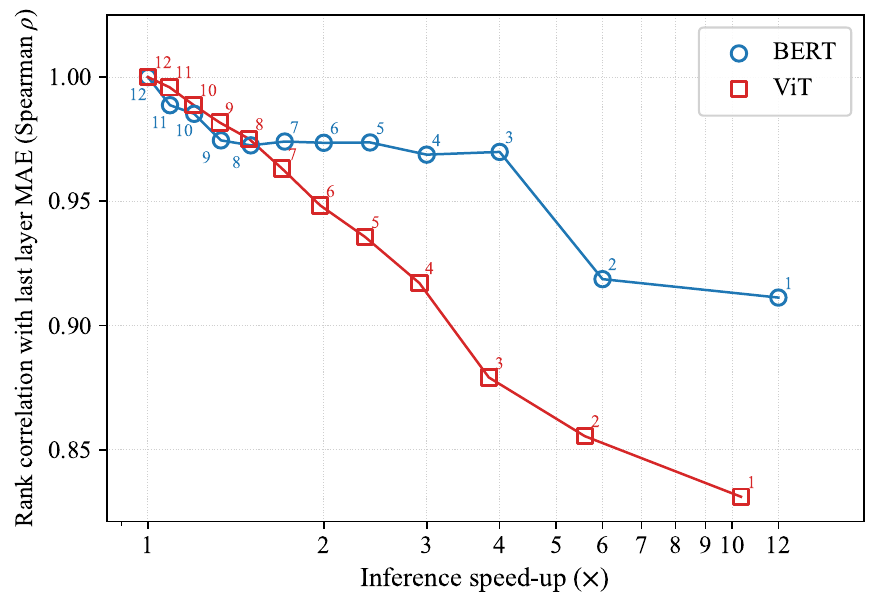}
    \end{subfigure}
    \hfill
    \begin{subfigure}[b]{0.235\textwidth}
        \centering
        \includegraphics[width=\textwidth]{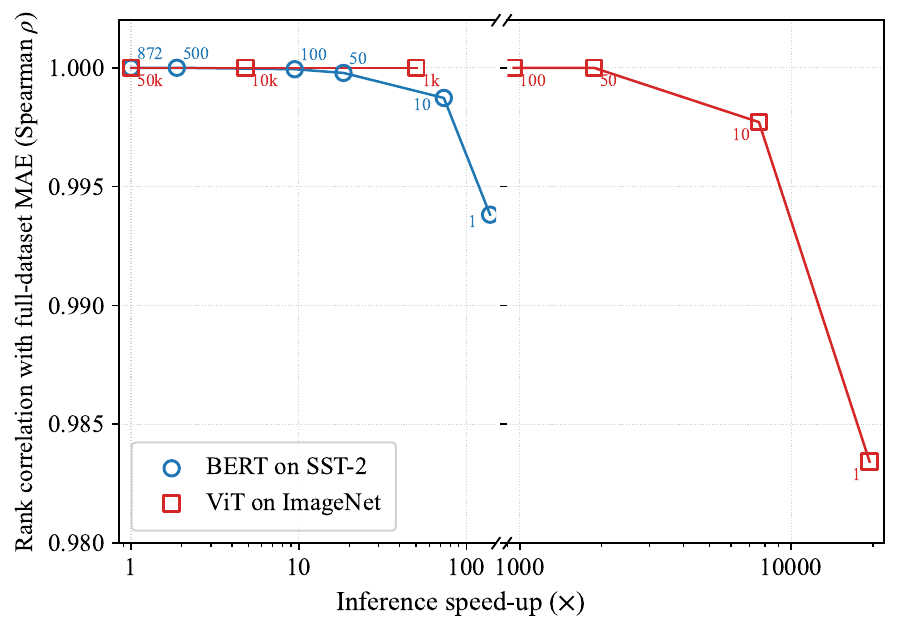}
    \end{subfigure}
    \\
    \begin{subfigure}[b]{0.235\textwidth}
        \centering
        \includegraphics[width=.9\textwidth]{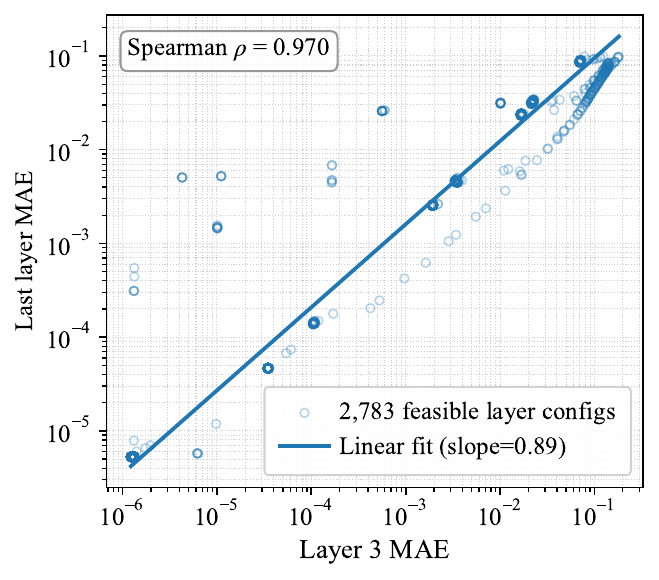}
    \end{subfigure}
    \hfill
    \begin{subfigure}[b]{0.235\textwidth}
        \centering
        \includegraphics[width=.9\textwidth]{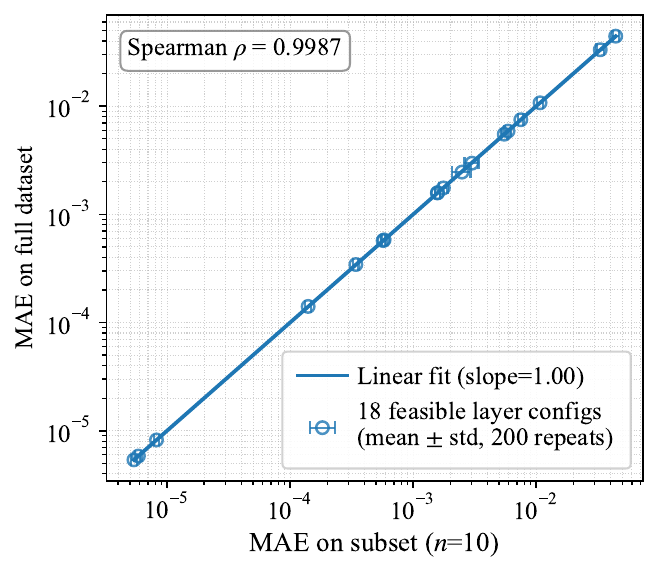}
    \end{subfigure}
    \\
    \begin{subfigure}[b]{0.235\textwidth}
    \centering
    \includegraphics[width=.9\textwidth]{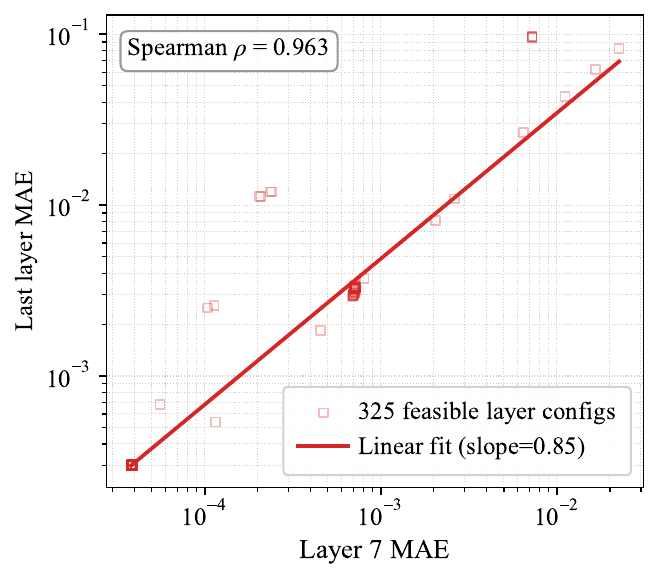}
    \end{subfigure}
    \hfill
    \begin{subfigure}[b]{0.235\textwidth}
        \centering
        \includegraphics[width=.9\textwidth]{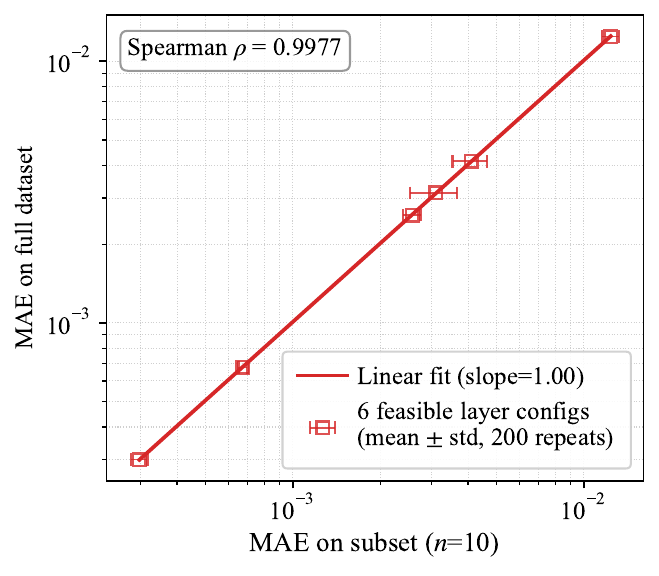}
    \end{subfigure}
    \caption{MAE at an early layer as a surrogate for MAE at the last layer, and the correlation between the MAE on the full and subset of the dataset, are shown on BERT and ViT.\label{fig:surr_layer}}
\end{figure}

\vspace{3pt}
\noindent\textbf{Data Subset Surrogate.} 
A single evaluation over the full dataset dominates the search time.
We observe that the MAE computed on a small random subset of the data is highly correlated with the full-dataset MAE.
For a given configuration, one can view the subset MAE as an estimate of the true full-dataset MAE.
That is, following Eq.~\eqref{eq:multi-obj}, we are not evaluating all $x \in \mathcal{D}$, but take a sampled subset of $\mathcal{D}$.
To assess the reliability of this surrogate, we measure the Spearman $\rho$ rank correlation between the two quantities: $\rho$ close to $1$ indicates that the ordering of configurations is largely preserved.
Figure~\ref{fig:surr_layer} (top right) reports $\rho$ and the resulting speed-up as a function of subset size.
For BERT, a subset of $10$ SST2 sentences achieves $\rho \approx 0.9987$, yielding a ${\sim}70\times$ speed-up over the full data, which is sufficient for reliable ranking.
For ViT, even $10$ ImageNet images yield $\rho \approx 0.9977$ with a ${\sim}7\text{K}\times$ speed-up.
Figure~\ref{fig:surr_layer} (middle right) and Figure~\ref{fig:surr_layer} (bottom right) illustrate this correlation for the chosen sizes: each point represents one configuration whose 
evaluation is performed on a random subset, with the corresponding full-dataset MAE on the vertical axis.

\vspace{3pt}
\noindent\textbf{Early-Layer Proxy.} 
In the \textsc{LayerProblem}, all layers share the same configuration; consequently, the MAE measured at intermediate Transformer layers is strongly correlated with the final-layer MAE.
Figure~\ref{fig:surr_layer} (top left) plots Spearman $\rho$ between early layer-$i$ MAE and last layer MAE against the resulting speed-up.
For BERT, layer~3 already gives $\rho \approx 0.97$ with ${\sim}4\times$ speed-up; for ViT, layer~7 achieves $\rho \approx 0.96$ with ${\sim}1.7\times$ speed-up.
Figure~\ref{fig:surr_layer} (middle left) and Figure~\ref{fig:surr_layer} (bottom left) show the corresponding relations between last layer MAE and layer-$i$ MAE.
Stopping the forward pass at an early layer avoids the computation of all subsequent Transformer layers, yielding an acceleration proportional to $L / i$.
This proxy is only valid when all layers are identical (Stage~1) and is disabled in Stage~2, where per-layer configurations vary independently.

For LLaMA3-8B, we use a data subset surrogate with 10 MMLU-dev sentences (due to its higher per‑batch GPU memory cost), achieving $\rho\approx0.996$ and ${\sim}28\times$ speed-up, along with a layer-3 MAE early proxy, which yields $\rho\approx0.727$ and ${\sim}10.7\times$ speed-up.

\vspace{3pt}
\noindent\textbf{Distributed Evaluation.} 
Evaluation is further accelerated by distributing the population across multiple GPUs, exploiting the parallelism of evolutionary search: the fitness of every individual in a generation can be computed independently.

\vspace{3pt}
\noindent\textbf{Overall Speed-Up.} 
Combining the above techniques reduces the effective cost of a single 
evaluation
by a factor of several hundred to over one thousand in Stage~1.
In Stage~2, we drop the early-layer proxy as different layers may use different configurations, yet the search is already warm-started in a promising region of the objective space.
Together, these techniques make the \methodname{} tractable even for large models such as LLaMA3-8B.

\section{Experimental Validation}
\methodname{} automatically approximates Transformers for efficient FHE inference through evolutionary search.
We evaluate it by asking, in turn, whether it works, how widely it applies, and what makes it work:
\begin{itemize}
    \item \textbf{RQ1:} ~\emph{Does \methodname{} discover approximation configurations that improve on the accuracy--depth and accuracy--latency trade-offs attained by hand-tuned FHE Transformer inference?} (Section~\ref{sec:acc-depth_pareto})
    \item \textbf{RQ2:} ~\emph{Does \methodname{} retain that advantage when applied to other Transformer architectures and deployed on other FHE backends?} (Section~\ref{sec:generalization})
    \item \textbf{RQ3:} ~\emph{What makes \methodname{} practical---which components of its two-stage search produce that advantage, and what does the search cost?} (Section~\ref{sec:searcheff})
\end{itemize}

\subsection{Experimental Setup}

\paragraph{Datasets.}
During search, BERT uses $10$ GLUE~\cite{wang2018glue} validation sentences as the proxy subset for corresponding tasks. For the evaluation, BERT is assessed on three GLUE tasks: SST-2, RTE, and QNLI.
The LLaMA uses $10$ random MMLU-dev~\cite{hendrycks2021measuring} samples for searching and is evaluated separately on three subsets of MMLU, with 64 samples drawn from each: abstract algebra, professional law, and clinical knowledge.
ViT uses $10$ random ImageNet-1K~\cite{deng2009imagenet} images as the search proxy, and the full validation set for Top-1 accuracy measurement.

\paragraph{Hardware and Libraries.}
The multi‑objective search is implemented on top of \texttt{PyMOO~0.6.1.6}, a Python framework for evolutionary optimization on cleartext polynomial model evaluation, with \texttt{HuggingFace} and \texttt{Transformers~4.57.1} providing off‑the‑shelf model architectures and weights, and our tensor backend is the \texttt{PyTorch~2.6.0} + \texttt{CUDA~12.4}.
To demonstrate that \methodname{} is backend-agnostic, we evaluate it on two distinct FHE backends: (1) \texttt{Phantom-FHE}~\cite{yang2024phantom}, for which we implemented our inference system in CUDA/C++ on top of it and NEXUS-EndtoEnd~\cite{nexuse2e}; and (2) \texttt{Desilo-FHE}~\cite{desilo}, which supports both CPU and GPU execution.
Our inference system additionally relies on \texttt{LibTorch~2.6.0} and \texttt{CUDA~12.4} for model weight loading.
All search runs and FHE inference latency measurements were conducted on a server equipped with the modified NVIDIA RTX 4090 48~GB GPUs and an AMD EPYC 7343 16-core 32-thread processor with 256~GB RAM.

\paragraph{Search Parameters.}
The two-stage NSGA-II shares a unified set of hyperparameters across all three architectures.
In \textsc{LayerProblem}, the population size is 48 and the algorithm runs for 50 generations, yielding $48 \cdot (50{+}1) = 2{,}448$ FEs.
In \textsc{NetworkProblem}, the population size is 96 and the algorithm runs for 225 generations, yielding $96 \cdot (225{+}1) = 21{,}696$ FEs.
The total FE budget per run is therefore $24{,}144$ across all models.
Both crossover and mutation operators are consistent across stages:
crossover is a two-point homogeneous operation with probability $0.9$;
mutation applies an integer step perturbation to each individual with probability $0.9$, while the per-variable mutation rate is set to $1/n_{\text{var}}$.

We use two objective metrics: MAE (defined in Eq.~\eqref{eq:multi-obj}) and the total approximation multiplicative depth.
The \emph{total approximation multiplicative depth} (denoted \emph{depth}) is the sum of the multiplicative depths of all non-linear approximation components in Transformers; the depth of homomorphic MMs is excluded because it is fixed for a given architecture and does not affect the search.

\paragraph{RNS‑CKKS Parameters.}
We follow the cryptographic parameterization as prior FHE inference works~\cite{lee2022low, ao2024autofhe, zhang2024secure}.
The cyclotomic polynomial degree is $N = 2^{16}$; the ciphertext budget is $D = 28$;
bootstrapping consumes $K = 14$ levels, leaving $D-K=14$
multiplicative levels after a refresh.
The base, special, and bootstrapping moduli each use $51$ bits, while the default modulus uses $46$ bits~\cite{cheon19ahybrid}.
This configuration guarantees $128$-bit security.

\paragraph{Bootstrapping Placement.}
Our search is depth-driven, since latency also depends on backend, bootstrapping placement, and hardware, which vary across deployments; and depth is set purely by the polynomial configurations and is packing- and backend-independent.
As a result, our bootstrapping placement is applied post-hoc under a fixed level budget, following a manual strategy similar to that of THOR~\cite{moon2025thor}: a level check is placed before every heavy operator, and bootstrapping is triggered only when the remaining level of a ciphertext is insufficient for the next operation.
Each bootstrapping acts on the $32{,}768$ slots, and the reported {\#B} is the number of such 32K-slot bootstrappings consumed by a full forward pass.
A better optimized bootstrapping placement strategy in future will yield further latency reduction.

\paragraph{Baselines.}
\textsc{Architecture Baseline:} We benchmark the proposed \methodname{} on three representative Transformer architectures with open-source weights in \texttt{HuggingFace}: the encoder of BERT‑base ($L{=}12$), the vision encoder of ViT‑Base ($L{=}12$), and the decoder of LLaMA3‑8B (${L=}32$).
For BERT and LLaMA, the number of tokens is set to $128$;
the original ViT-Base (Patch16-224) sequence length of $197$ is padded to $256$ to match the slots.

\textsc{Approximation Baseline:} For comparison, we reproduce the non‑linear approximation recipes of NEXUS~\cite{zhang2024secure} and THOR~\cite{moon2025thor} within our FHE inference system.
For the iterative softmax of Cho~et~al.~\cite{cho2024fast}, we pair it with
Chebyshev invsqrt~\cite{badawi2026sok, yang2025arion} in LayerNorm/RMSNorm and a Chebyshev GELU/SiLU~\cite{ebel2025orion, yang2025arion} of degree $511$ to obtain a high-precision complete end‑to‑end pipeline.
All baselines are evaluated under the same cryptographic and hardware settings.
Because the level of input ciphertext, homomorphic MM, packing strategies, and bootstrapping placements vary substantially across frameworks, our reproducing end‑to‑end latencies of NEXUS and THOR reported here reflect only the approximation configuration and are not directly comparable to those in the original papers;
we adopt a row‑packing homomorphic MM (discussed in Section~\ref{sec:fhe_system}) to ensure a fair, isolated evaluation of the approximation recipes alone.

\begin{figure*}[t]
    \centering
    \begin{subfigure}[b]{0.325\textwidth}
        \centering
        \includegraphics[width=0.985\textwidth]{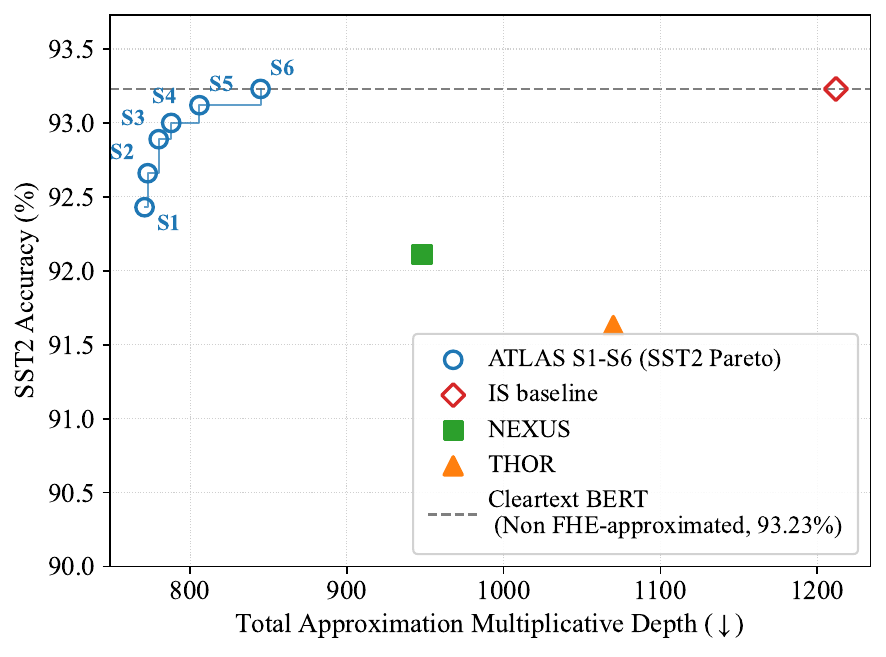}
        \caption{SST-2}\label{fig:fig_bert_pareto_sst2}
    \end{subfigure}
    \hfill
    \begin{subfigure}[b]{0.325\textwidth}
        \centering
        \includegraphics[width=0.985\textwidth]{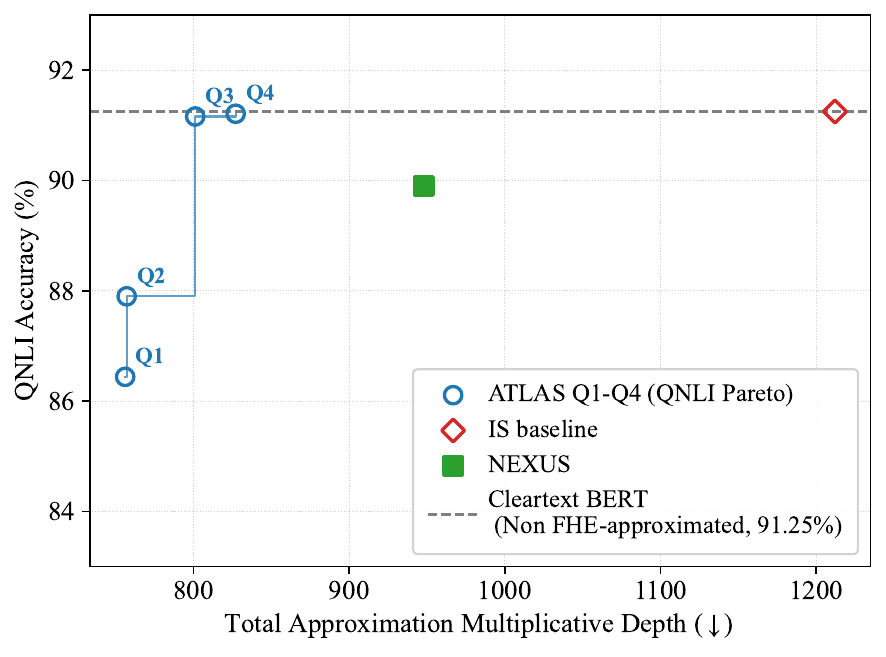}
        \caption{QNLI}\label{fig:fig_bert_pareto_qnli}
    \end{subfigure}
    \hfill
    \begin{subfigure}[b]{0.325\textwidth}
        \centering
        \includegraphics[width=0.985\textwidth]{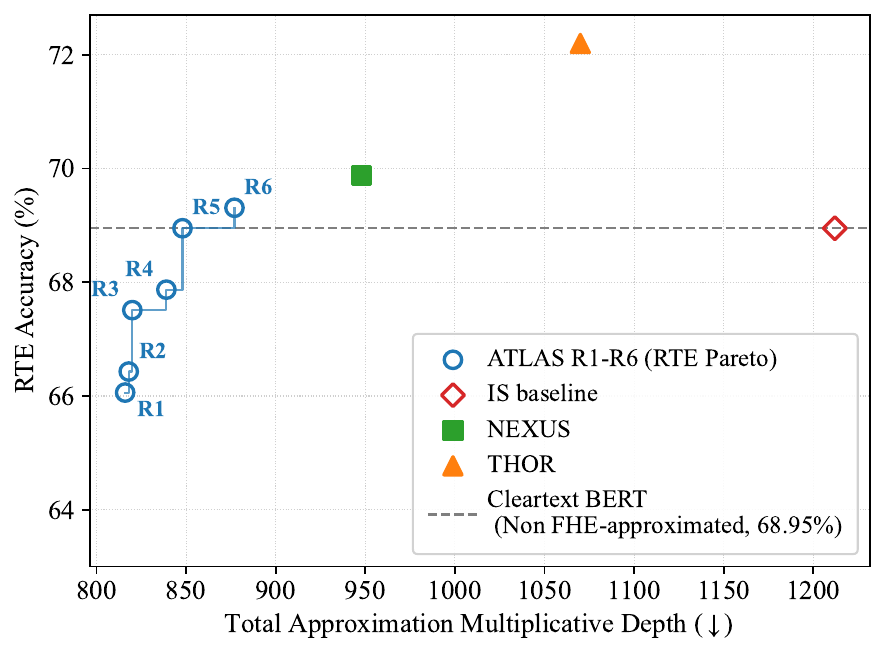}
        \caption{RTE}\label{fig:fig_bert_pareto_rte}
    \end{subfigure}
        \caption{Pareto fronts discovered by \methodname{} on three tasks of BERT. Each point represents a non‑dominated $\bm{\lambda}$ configuration, plotted in the space of multiplicative depth vs.\ downstream task accuracy. The iterative softmax baseline is shown as a reference. For NEXUS and THOR, accuracy is as reported in the respective original papers, whereas depth is measured in our inference system. Dashed horizontal lines indicate the accuracy of the cleartext, non FHE-approximated model. The corresponding accuracy--latency Pareto fronts, measured end-to-end under FHE, are reported in Figure~\ref{fig:bert_acc_vs_lat} of Appendix~\ref{sec:appendix_lat}.}\label{fig:pareto_front_bert_depth}
\end{figure*}

\begin{table*}[t]
\centering
\caption{Latency of SST2 BERT Pareto-optimal configurations under different FHE backends. Depth -- total approximation multiplicative depth; \#B -- the number of bootstrapping; Lat. -- FHE inference latency; Lat. Redu. -- latency reduction rate; MAE -- mean absolute error of the FHE output logits with respect to the cleartext model.}
\label{tab:pareto_lat_combined}
\resizebox{0.9\textwidth}{!}{%
\begin{tabular}{l c cccc cccc cccc}
\toprule
\multirow{2}{*}{\makecell[c]{Config}} & \multirow{2}{*}{\makecell[c]{Mul.\\Depth}} & \multicolumn{4}{c}{\texttt{Phantom-FHE} (GPU)} & \multicolumn{4}{c}{\texttt{Desilo-FHE} (GPU)} & \multicolumn{4}{c}{\texttt{Desilo-FHE} (CPU)} \\
\cmidrule(lr){3-6} \cmidrule(lr){7-10} \cmidrule(lr){11-14}
 & & \#B & \makecell[c]{Lat. (s)} & \makecell[c]{Lat. Redu.} & \makecell[c]{MAE} & \#B & \makecell[c]{Lat. (s)} & \makecell[c]{Lat. Redu.} & \makecell[c]{MAE} & \#B & \makecell[c]{Lat. (s)} & \makecell[c]{Lat. Redu.} & \makecell[c]{MAE} \\
\midrule
IS baseline & 1212 & 600 & 1054 & - & 0.007 & 603 & 489 & - & 0.008 & 603 & 12067 & - & 0.009 \\
\midrule
\methodname{}-S1 & 771 & 372 & $\bf{779}$ & $\bf{26.1\%}$ & 0.188 & 371 & $\bf{288}$ & $\bf{41.1\%}$ & 0.021 & 371 & 6814 & 43.5\% & 0.021 \\
\methodname{}-S2 & 773 & 372 & 784 & 25.6\% & 0.105 & 371 & 291 & 40.5\% & 0.015 & 371 & $\bf{6579}$ & $\bf{45.5\%}$ & 0.015 \\
\methodname{}-S3 & 780 & 372 & 780 & 26.0\% & 0.099 & 375 & 289 & 40.9\% & 0.010 & 375 & 6693 & 44.5\% & 0.010 \\
\methodname{}-S4 & 788 & 375 & 787 & 25.3\% & 0.082 & 379 & 295 & 39.7\% & 0.012 & 379 & 6826 & 43.4\% & 0.012 \\
\methodname{}-S5 & 806 & 396 & 809 & 23.2\% & 0.050 & 383 & 301 & 38.4\% & 0.009 & 383 & 7056 & 41.5\% & 0.009 \\
\methodname{}-S6 & 845 & 399 & 809 & 23.2\% & 0.009 & 379 & 315 & 35.6\% & 0.011 & 379 & 7413 & 38.6\% & 0.011 \\
\bottomrule
\end{tabular}%
}
\end{table*}

\subsection{Results of \methodname{} \label{sec:main-results}}
For each neural architecture, \methodname{} obtains a set of MAE--depth Pareto configurations with two-stage search (as shown in Figure~\ref{fig:atlas_framework}) on a small set of proxy data from tasks.
Then, \methodname{} evaluates these configurations on each downstream task to build the accuracy--depth Pareto front.
The final accuracy is reported on the full validation set, while the search uses only small non-overlapping samples to avoid overfitting.

\subsubsection{Main Results on BERT}\label{sec:acc-depth_pareto}

\paragraph{Comparison on Accuracy--Latency/Depth Trade-off.}\label{sec:bert_acc-depth_pareto}
The trade-off between predictive accuracy and inference cost is evident in the Pareto fronts of Figure~\ref{fig:pareto_front_bert_depth} and the corresponding measurements in Table~\ref{tab:pareto_lat_combined}: no configuration is simultaneously the most accurate and the least expensive.
Its shape, however, is far from fixed: it varies markedly from one task to the next.
The three fronts of Figure~\ref{fig:pareto_front_bert_depth} span comparable cost ranges ($61$--$74$ levels of depth and $12$--$29$~s of latency), yet the accuracy relinquished across that range differs by a factor of six: $0.80$ pp on SST-2, $3.25$ pp on RTE, and $4.77$ pp on QNLI.
A single accuracy tolerance therefore cannot serve all three tasks: a $1$ pp budget admits every configuration on SST-2 ($6$ of $6$), but only $2$ of $4$ on QNLI and $2$ of $6$ on RTE.
This observation motivates the multi-objective formulation of Eq.~\eqref{eq:multi-obj} in preference to the thresholded form of Eq.~\eqref{eq:orig_prac}: a tolerance $\epsilon$ calibrated on one task is miscalibrated on the next, whereas a Pareto front permits the operating point to be selected after the search rather than fixed beforehand.
Our hand-constructed IS baseline reaches the accuracy of the cleartext, non-approximated BERT on all three tasks ($93.23\%$, $91.25\%$, and $68.95\%$), and \methodname{} matches it on SST-2 and RTE and comes within $0.04$ pp on QNLI---but at markedly different cost.
At matched accuracy, \methodname{} reduces multiplicative depth by $30.7\%$ on average ($30.3\%$, $31.8\%$, and $30.0\%$) and end-to-end latency by $36.1\%$ ($35.6\%$, $35.9\%$, and $36.7\%$), the SST-2 case being reported in full in Table~\ref{tab:pareto_lat_combined}.
At the high-accuracy end of the front, S6 matches cleartext SST-2 exactly ($93.23\%$) at $30.3\%$ lower depth and $35.6\%$ lower latency, while R6 marginally exceeds cleartext RTE ($69.31\%$ against $68.95\%$) at $27.6\%$ lower depth and $36.1\%$ lower latency; the latter margin corresponds to one additional correct prediction out of $277$ and is therefore best interpreted as parity.
The preferred configuration likewise differs by task---S6, Q4, and R6---a task-specific preference that no single hand-tuned configuration can express.
For \textbf{RQ1}, \methodname{} thus attains hand-tuned accuracy at approximately two thirds of its depth and latency on encoder-only architectures, without manual tuning.

\paragraph{Comparison with NEXUS and THOR.}
We re-implement the non-linear approximations of NEXUS and THOR within our inference system, retaining the approximation algorithms and hyperparameters they specify for softmax, normalization, and activation---polynomial degrees and iteration counts---but substituting our level budget and bootstrapping placement.
NEXUS then operates at depth $948$ (\#B$=912$, $1134$~s) and THOR at depth $1070$ (\#B$=732$, $1023$~s); all six \methodname{} configurations in Table~\ref{tab:pareto_lat_combined} require lower multiplicative depth, fewer bootstrapping operations, and less end-to-end latency than either.
The comparison is indicative rather than rigorous: packing, homomorphic MM, and bootstrapping placement are ours, so these measurements characterize the approximation hyperparameters alone and not the performance of the original systems, and the accuracies in Figure~\ref{fig:pareto_front_bert_depth} are as reported in the respective papers.
On SST-2 \methodname{} dominates both baselines on both axes, and on QNLI it dominates NEXUS, the only one of the two reporting that task; on RTE both report higher accuracy at greater depth, THOR's obtained with fine-tuned weights that are not publicly released.

\paragraph{Latency Breakdown.}
Figure~\ref{fig:time_breakdown} decomposes the latency of the IS baseline and \methodname{}-S1 in our inference system.
Non-linear operations and their bootstrapping account for $578$~s of the baseline's $1054$~s ($55\%$); \methodname{}-S1 cuts this by $48\%$ to $300$~s, essentially the whole of the $26\%$ end-to-end speedup, two thirds of it from bootstrapping and one third from cheaper polynomial evaluation.
Matrix multiplication is unchanged at ${\sim}460$~s, so these savings are additive with respect to the packing and matrix-multiplication optimizations pursued in parallel work (Section~\ref{sec:fhe_system}).

\subsubsection{Generalization to FHE Backend and Transformer Architecture}\label{sec:generalization}

\paragraph{Generalization across FHE Backends.}
The configurations in Table~\ref{tab:pareto_lat_combined} come from a single cleartext search against depth and are deployed unmodified on every backend.
Because depth is determined solely by the decision vector $\bm{\lambda}$, the relative depth reduction is invariant ($30.3\%$--$36.4\%$); the relative latency reduction is not, attaining $23.2\%$--$26.1\%$ on our \texttt{Phantom-FHE} system, $35.6\%$--$41.1\%$ on \texttt{Desilo-FHE} GPU, and $38.6\%$--$45.5\%$ on \texttt{Desilo-FHE} CPU---conversion ratios of approximately $0.72$, $1.14$, and $1.24$.
Our system converts least efficiently because its unoptimized homomorphic MM accounts for ${\sim}460$~s of the $1054$~s baseline; on the faster backends the ratio exceeds unity, so the latency reduction attainable by \methodname{} increases as the linear components are optimized.
Transfer also holds across tasks: on \texttt{Desilo-FHE} GPU, the QNLI and RTE configurations reduce latency by $36\%$--$42\%$ and $36\%$--$39\%$ (Appendix~\ref{sec:appendix_lat}).

\begin{figure}[t]
    \centering
    \begin{subfigure}[b]{0.235\textwidth}
        \centering
        \includegraphics[width=\textwidth]{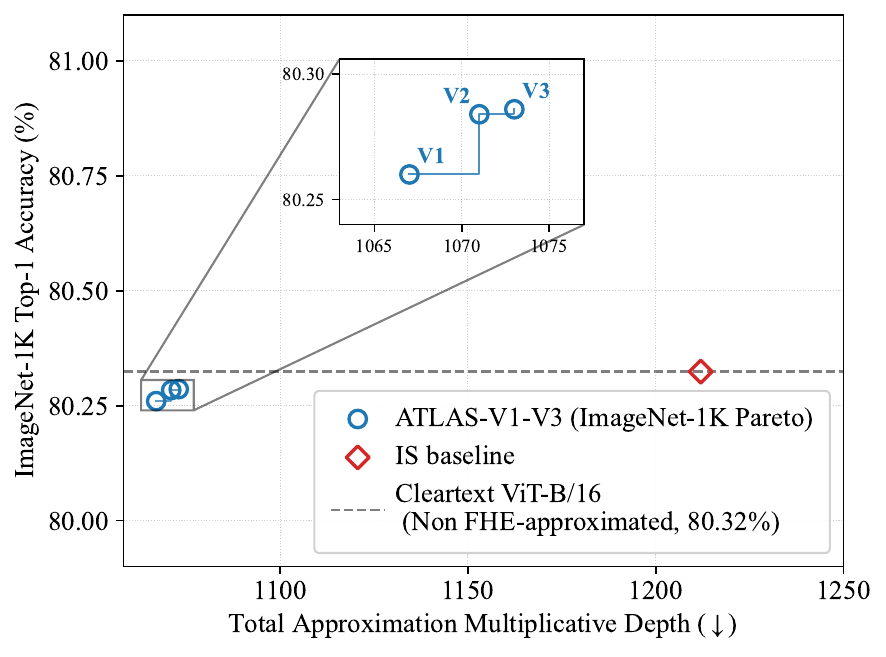}
        \caption{ImageNet-1K}\label{fig:fig_vit_pareto_depth_imagenet}
    \end{subfigure}
    \hfill
    \begin{subfigure}[b]{0.235\textwidth}
        \centering
        \includegraphics[width=\textwidth]{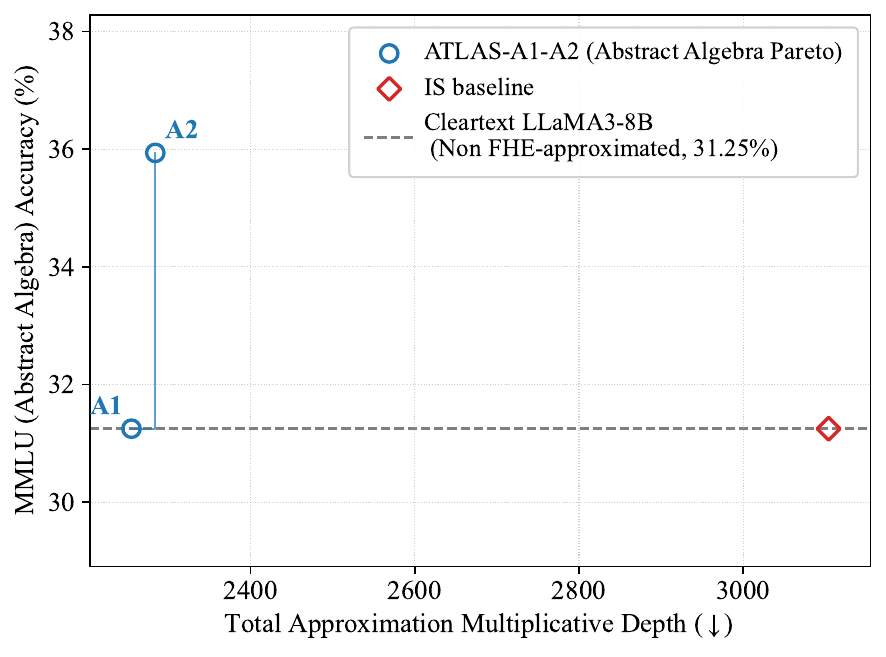}
        \caption{MMLU Abstract Algebra}\label{fig:fig_llama_pareto_mmlu_aa}
    \end{subfigure}
    \\
    \begin{subfigure}[b]{0.235\textwidth}
        \centering
        \includegraphics[width=\textwidth]{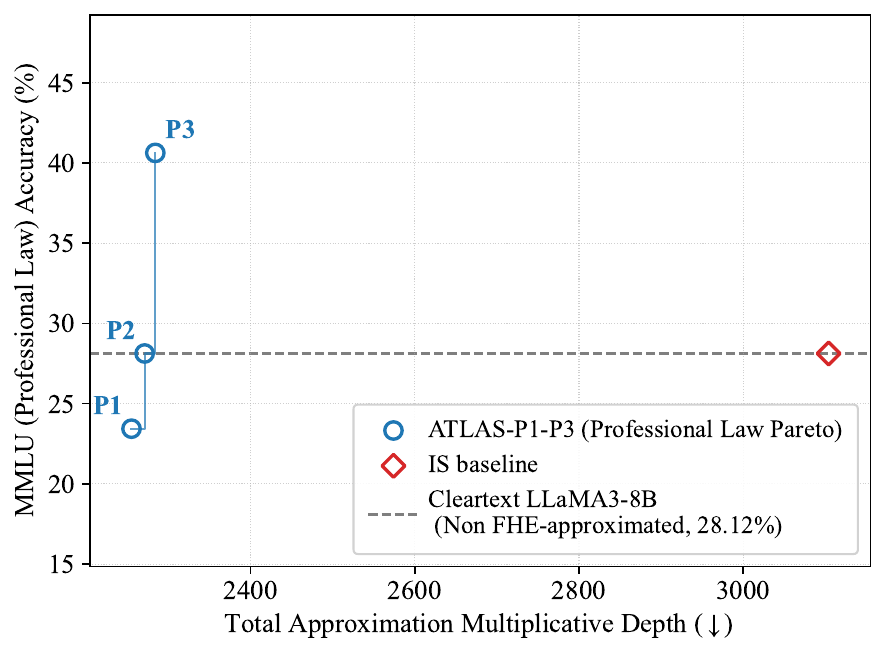}
        \caption{MMLU Professional Law}\label{fig:fig_llama_pareto_mmlu_pl}
    \end{subfigure}
    \hfill
    \begin{subfigure}[b]{0.235\textwidth}
        \centering
        \includegraphics[width=\textwidth]{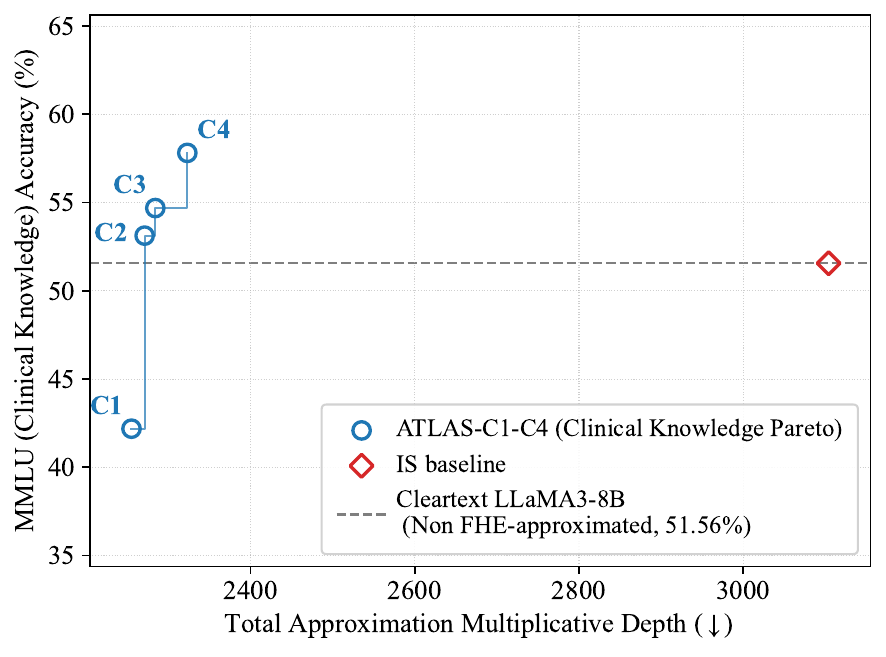}
        \caption{MMLU Clinical Knowledge}\label{fig:fig_llama_pareto_mmlu_ck}
    \end{subfigure}
    \caption{
    Pareto fronts discovered by \methodname{} on ImageNet-1K of ViT and three MMLU tasks of LLaMA.
    }\label{fig:pareto_front_llama}
\end{figure}

\paragraph{Generalization across Transformer Architectures.}
The same procedure, again unmodified, transfers to a vision encoder and a decoder-only LLM.
On ImageNet-1K (Figure~\ref{fig:fig_vit_pareto_depth_imagenet}), \methodname{}-V3 matches the IS baseline to within $0.03$ pp ($80.29\%$ against $80.32\%$) at $11.5\%$ lower depth, giving $2259$~s against $2654$~s ($14.9\%$) and reducing bootstrappings from $1728$ to $1362$ (Appendix~\ref{sec:appendix_lat}).
ViT nonetheless tolerates far less approximation than BERT, which retains cleartext accuracy at $30\%$ lower depth, and its three non-dominated configurations span only six levels ($1067$--$1073$).
For LLaMA3-8B, \methodname{} reduces depth by $25\%$--$27\%$ from the baseline's $3104$ across three MMLU subsets (Figures~\ref{fig:fig_llama_pareto_mmlu_aa}--\ref{fig:fig_llama_pareto_mmlu_ck}) at parity in accuracy.
Evaluation is zero-shot and no-CoT, truncated to $128$ tokens on $64$ examples per subset; the truncation lowers absolute accuracy on the longer-form subsets, and one answer corresponds to $1.6$ pp, so we read the observed differences as parity rather than gains.
The three fronts largely coincide (A1$=$P1$=$C1, P2$=$C2, A2$=$P3$=$C3), indicating similar approximation requirements across subsets.
We report depth rather than latency at this scale, as an end-to-end FHE pass of a $32$-layer $8$B-parameter decoder exceeds our single-GPU budget and depth is, as shown above, backend-independent.
For \textbf{RQ2}, \methodname{} therefore transfers across encoder-only, vision, and decoder-only architectures, though the attainable depth reduction differs substantially among them.

\subsection{Ablative Analysis of \methodname{}}\label{sec:searcheff}
We now isolate the contributions of the two components that make the search practical, in the order they were introduced: the two-stage joint optimization strategy of Section~\ref{sec:twostage_search} and the acceleration techniques of Section~\ref{sec:accelerate}.
All ablations are conducted on BERT.

\subsubsection{Impact of Optimization Strategies}
\begin{figure}[t]
    \centering
    \begin{subfigure}[b]{0.235\textwidth}
        \centering
        \includegraphics[width=\textwidth]{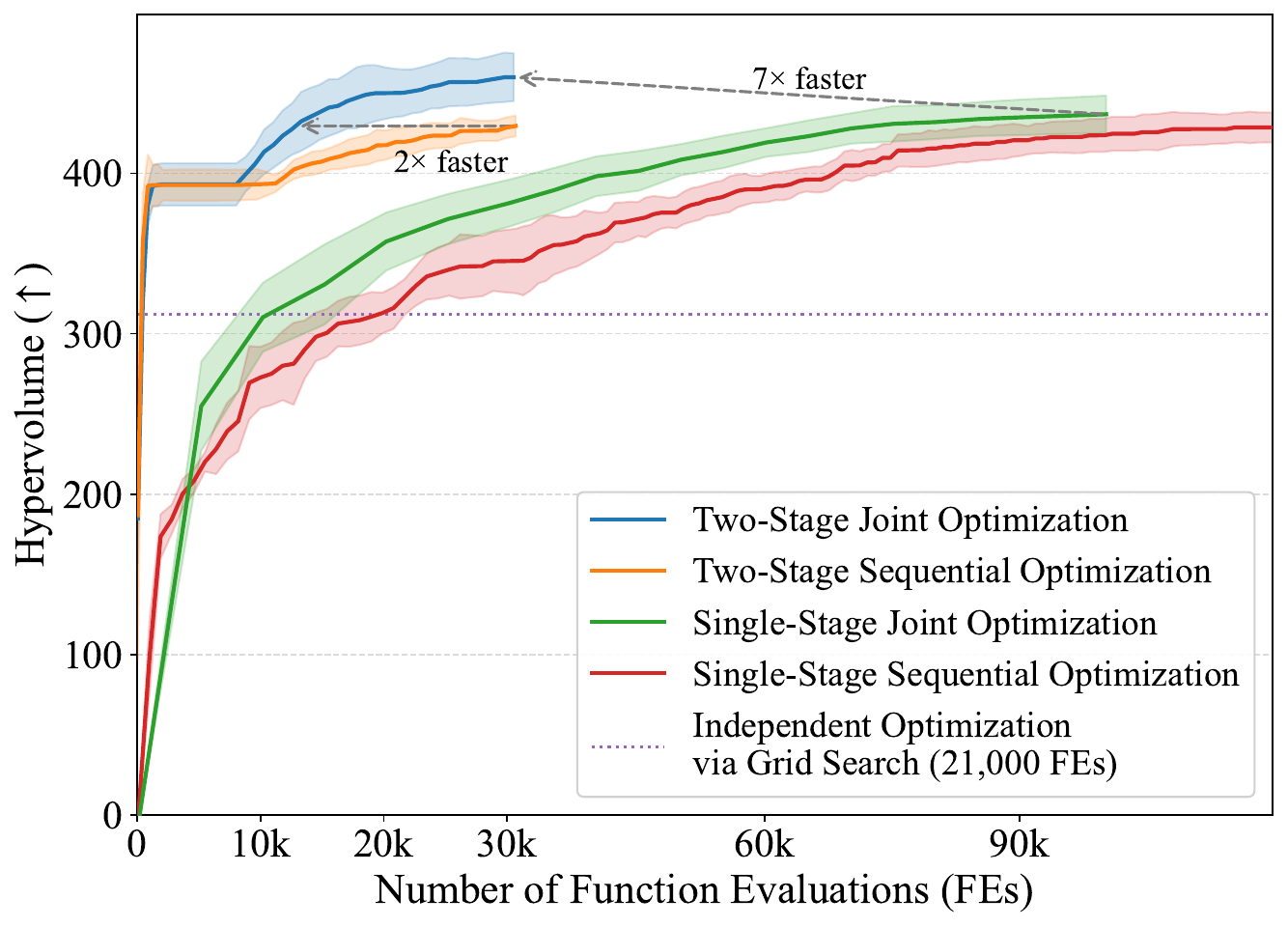}
        \caption{HV vs. number of FEs}\label{fig:searcheff_opt}
    \end{subfigure}
    \hfill
    \begin{subfigure}[b]{0.235\textwidth}
        \centering
        \includegraphics[width=\textwidth]{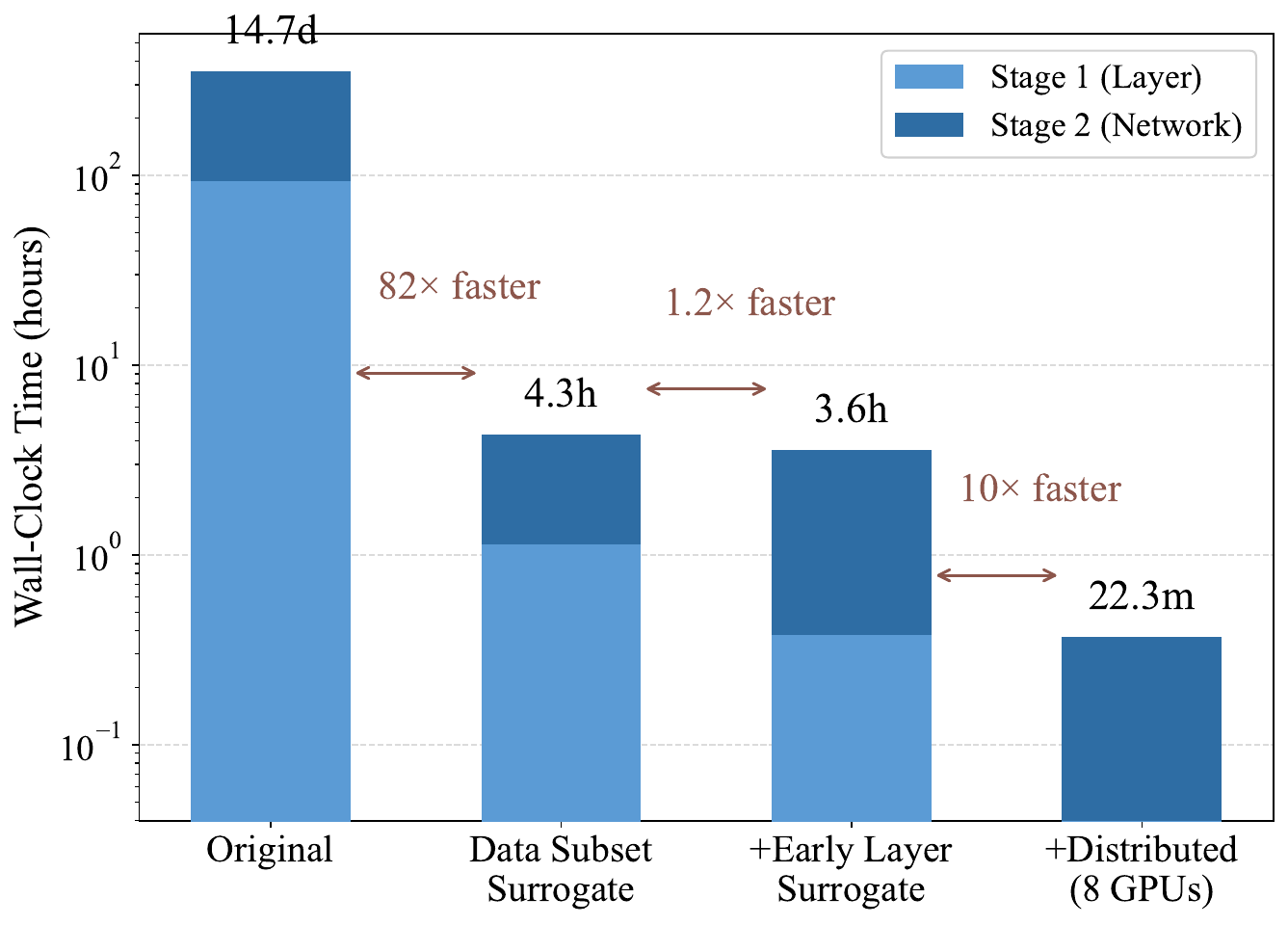}
        \caption{HV vs. wall-clock time}\label{fig:searcheff_accel}
    \end{subfigure}
    \\
    \begin{subfigure}[b]{0.235\textwidth}
        \centering
        \includegraphics[width=\textwidth]{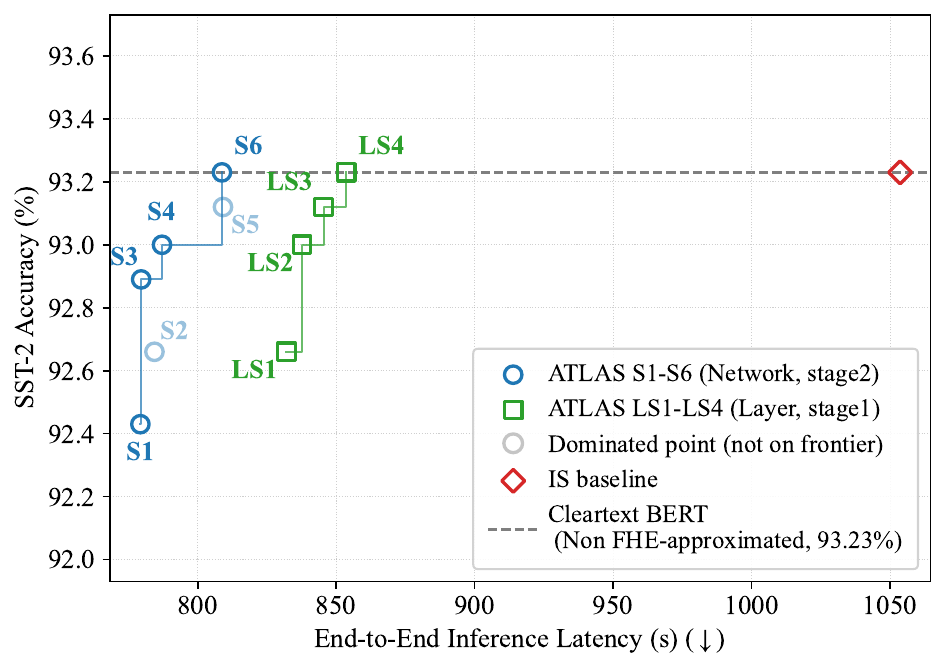}
        \caption{Stage 1 vs. 2, \texttt{Phantom-FHE}}\label{fig:fig_stage1_vs_stage2_atlas}
    \end{subfigure}
    \hfill
    \begin{subfigure}[b]{0.235\textwidth}
        \centering
        \includegraphics[width=\textwidth]{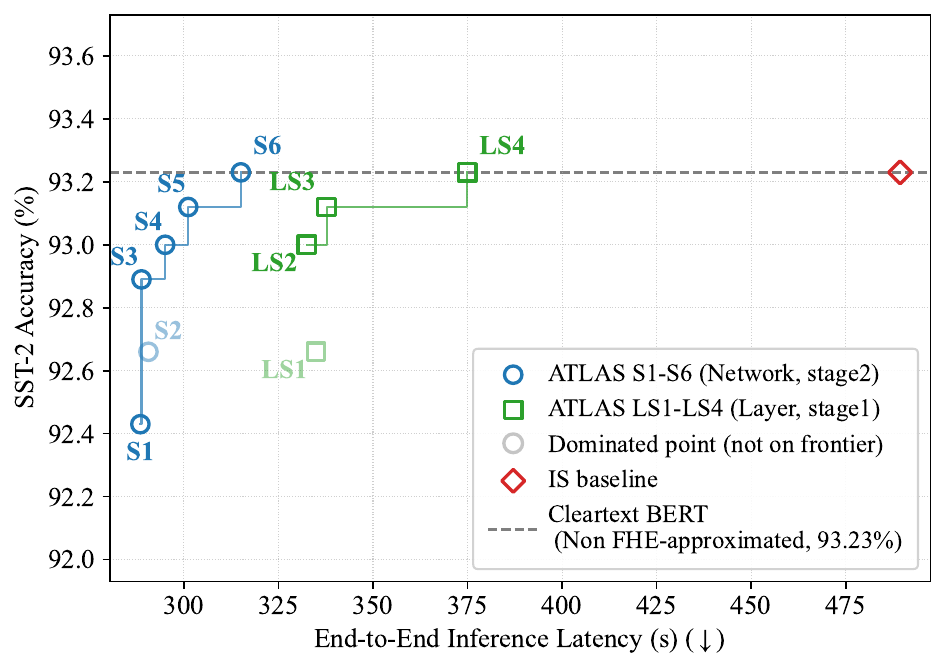}
        \caption{Stage 1 vs. 2, \texttt{Desilo-FHE}}\label{fig:fig_stage1_vs_stage2_desilo}
    \end{subfigure}
    \caption{Ablative analysis of \methodname{} on BERT.
             (a) Comparison of optimization strategies, in terms of hypervolume reached for a given number of function evaluations.
             (b) Impact of the acceleration techniques of Section~\ref{sec:accelerate}: each curve adds one technique on top of the previous.
             (c,~d) Latency of the Stage~1 (LS1--LS4) and Stage~2 (S1--S6) Pareto configurations on BERT/SST-2, under the two FHE backends.}
    \label{fig:ablation}
\end{figure}

Figure~\ref{fig:searcheff_opt} tracks the HV of the Pareto front against the number of FEs for five strategies; recall that a larger HV in the (depth, MAE) space indicates a better trade-off between multiplicative depth and approximation error.
As a lower bound, independent optimization via grid search---separately gridding the softmax space and the combined normalization and activation space, then taking the best combination---requires roughly $21{,}000$ FEs to reach an HV of ${\sim}310$.

\paragraph{Sequential vs.\ Joint Optimization.}
Sequential optimization configures one operator group at a time while freezing the others, cycling through softmax, normalization, and activation; joint optimization evolves all variables together.
Joint optimization dominates at both stage settings: two-stage joint reaches an HV of ${\sim}420$ in ${\sim}14{,}000$ FEs, roughly twice as fast as its sequential counterpart, and single-stage joint stays above single-stage sequential over the whole run.
The three approximation blocks are therefore interdependent and cannot be tuned in isolation.

\paragraph{Single-stage vs.\ Two-stage Optimization.}
Two-stage joint optimization, our configuration, attains an HV of ${\sim}450$, whereas single-stage joint optimization plateaus at ${\sim}435$ despite consuming several times more evaluations.
At that plateau level the warm-start reaches the same HV in ${\sim}14{,}000$ FEs against ${\sim}100{,}000$, a $7\times$ reduction.
Turning from search progress to the configurations themselves, Figures~\ref{fig:fig_stage1_vs_stage2_atlas} and~\ref{fig:fig_stage1_vs_stage2_desilo} attribute this advantage to the division of labor between the stages.
Stage~1 searches the uniform space, which alone spans $10^7$ combinations, and converges from the baseline to an initial front (LS1--LS4); Stage~2 then refines these configurations per layer, pushing the front further left to S1--S6.
On BERT/SST-2 with the \texttt{Desilo-FHE} backend, Stage~1 accounts on average for about $75\%$ of the total latency reduction and Stage~2 for the remaining $25\%$, rising to ${\sim}35\%$ at S6, the configuration that recovers cleartext accuracy.
For \textbf{RQ3}, the two stages are therefore complementary: a user may stop after Stage~1 for a strong uniform configuration, or run Stage~2 for the additional per-layer refinement, whereas manual tuning is infeasible over $10^7$ combinations whose best choice varies by model and task.

\subsubsection{Impact of Acceleration Techniques}
Figure~\ref{fig:searcheff_accel} breaks down the wall-clock time required to reach a target HV as the acceleration techniques are incrementally enabled.
Two-stage NSGA-II without any surrogate would require roughly $14.7$ days to exhaust the search budget.
The data subset surrogate alone reduces this to $4.3$ hours, an $82\times$ speed-up, by evaluating MAE on $10$ representative samples rather than the full calibration set.
Adding the early-layer proxy in Stage~1 shortens the run to $3.6$ hours ($1.2\times$ further), since the forward pass is truncated at an intermediate layer whose MAE is strongly correlated with the last-layer MAE.
Distributed evaluation over $8$ GPUs brings the search down to $22.3$ minutes, a further $10\times$ from the parallelism inherent to population-based search.
Of those $22.3$ minutes, Stage~1 accounts for roughly two, so the uniform configuration is available almost immediately and the remainder is spent on per-layer refinement.
For \textbf{RQ3}, these techniques together reduce the search on BERT from more than two weeks to under half an hour, a factor of ${\sim}950$, bringing automated per-layer configuration within a routine compute budget.

\section{Related Work}

\subsection{Non-linear Approximation Design}
\textbf{Designing New Approximations.}
To enable secure Transformer inference under FHE, recent efforts focus on designing approximations for the key non-linear components: softmax, normalization, and activation.
One line of works~\cite{zimerman2024converting, park-etal-2025-powerformer, rho2025encryption} design HE-friendly alternatives and retrain the entire model; however, full retraining or fine-tuning is computationally intensive and expensive for modern LLMs.
The most relevant direction to our work is therefore the post-training approach: replacing the non-linear operations directly with high-precision polynomial approximations while keeping the pretrained weights frozen, as done in NEXUS~\cite{zhang2024secure}, THOR~\cite{moon2025thor}, and MOAI~\cite{zhang2025moai}, etc.
Additionally, several studies specifically tackle the most challenging operator in Transformer, softmax, with specialized approximation strategy~\cite{cho2024fast, moon2025thor, park2026efficient}.

\noindent\textbf{Manual Configuration.}
Despite their effectiveness, all above methods rely on heuristics and expert experience to set approximation hyperparameters such as polynomial degree and iteration counts.
These approximations are often adapted from numerical methods (e.g., Goldschmidt algorithm~\cite{goldschmidt1964applications}, Newton's method, and Remez algorithm for power-basis/Chebyshev polynomial, etc.), and a significant safety margin is deliberately built in to ensure correctness.
Within deep neural networks, however, the precision requirements typically leave a considerable slack that heuristic or hand-tuned configurations often fail to exploit.

\noindent\textbf{Automatic Configuration.}
The work closest to ours is AutoFHE~\cite{ao2024autofhe}, which uses multi-objective optimization and full fine-tuning to search per-layer ReLU approximations in FHE CNNs from a small set of three candidate degrees (identity, quadratic, and minimax composite~\cite{lee2021minimax, lee2023precise}).
However, this approach is limited to CNNs and does not readily transfer to Transformers, especially modern LLMs. Since the number of non-linearities is much larger, the search space explodes, the functional forms are more diverse, and the fine-tuning cost becomes prohibitive.
More recently, OLA~\cite{lee26ola} and SLOTHE~\cite{man26slothe} have explored layer-specific polynomial approximations for non-linearities—OLA for CNNs via degree selection, and SLOTHE for certain non-linearities such as GELU in Transformers via code decomposition—but neither addresses the automated configuration of \emph{softmax, normalization, and activation functions jointly} in Transformers.
Prior to AutoFHE, several works applied neural architecture search to FHE CNNs to reduce the number of approximated ReLUs and select architectures~\cite{srinivasan2019delphi, ghodsi2020cryptonas, jha2021deepreduce}.
These methods require neural architectural modifications and training from scratch, making them equally ill-suited for Transformers.

In contrast, \methodname{} is the first method to automatically configure per-layer polynomial approximations for FHE Transformers. Unlike AutoFHE, whose search is coupled to gradient-based fine-tuning of the network, \methodname{} operates entirely post-training on a frozen pretrained model, and therefore scales to settings where fine-tuning is prohibitively expensive, such as the 8B-parameter LLaMA3. It further generalizes across encoder-only (BERT), decoder-only (LLaMA3), and vision (ViT) Transformers---rather than the ReLU-only CNNs targeted by prior automated methods---systematically exploiting the precision slack that hand-tuned configurations leave unused.

\subsection{FHE Inference System Design}\label{sec:fhe_system}

\noindent\textbf{Efficient Homomorphic Matrix Multiplication.}
Several recent frameworks have focused on optimizing packing and homomorphic MM to reduce overhead.
For example, NEXUS~\cite{zhang2024secure} eliminates wasted ciphertext slots through compression/decompression and exploits monomial properties to lower communication.
THOR~\cite{moon2025thor} formulates MM via diagonals, using ciphertext rotations and entrywise products to compute matrix products efficiently, and packs multiple diagonals into one ciphertext for parallelism.

\noindent\textbf{End-to-End Transformer Inference.}
While many works provide individual homomorphic operators, few deliver a complete, reproducible end-to-end pipeline.
NEXUS~\cite{zhang2024secure} offers standalone components, but how multiple packing strategies should coexist or interconvert across modules remains unclear.
THOR~\cite{moon2025thor} presents an end-to-end system, yet it requires precomputing and storing over 180~GB of model weights and keys, and demands at least 80~GB of GPU memory for a single BERT inference; moreover, the supplied configuration is tied to its precomputed weights and cannot easily generalize to other downstream tasks.
MOAI~\cite{zhang2025moai} and ARION~\cite{yang2025arion} both achieve end-to-end inference, but at substantial cost: MOAI reports its end-to-end evaluation on a high-memory datacenter GPU (an H200), whereas ARION runs on CPU only.
Also, their unamortized latency with batch size 256 reaches 10 and 16 hours per 128-token sequence, respectively, making the cost of deployment testing prohibitively high.
Although their amortized throughput is competitive, the underlying packing algorithms do not support smaller batch sizes without substantial rework.

Given these barriers, we pivot to a basic row-packing implementation built upon and extended from an open-source codebase NEXUS-EndtoEnd~\cite{nexuse2e}.
While the simple row-packing homomorphic MM is not as competitive as the highly optimized alternatives, it does not affect the configuration of non-linear approximations and provides a fair, transparent baseline for evaluating different approximation choices.

Our system supports the full lifecycle from approximation search to end-to-end ciphertext model evaluation.
Importantly, our approach is orthogonal to the design of efficient packing and MM; our search framework serves as a general-purpose tool that complements any of the above high-performance systems by providing automatic approximation configurations.

\section{Concluding Remarks}

The central insight of this paper is simple: existing FHE Transformer frameworks fix approximation hyperparameters per function type and apply them uniformly across layers, leaving significant depth savings unrealized. \methodname{} addresses this by treating approximation configuration as an optimization problem rather than a design choice, automatically discovering per-layer configurations that existing frameworks can adopt without modification. Across three Transformer architectures and two FHE backends, \methodname{} delivers a 35 percent reduction in multiplicative depth and a 35 percent reduction in inference latency, with negligible accuracy loss and no manual effort, in under one hour of search.

We also demonstrate that the approach extends to vision Transformers, to our knowledge the first automated per-layer approximation configuration for that family.
Vision carries much of the practical demand for encrypted inference: medical imaging, biometrics, and related image-based diagnostics are among the settings in which a client is least able to disclose its input, and vision Transformers have displaced the convolutional models that earlier private-inference work targeted.

\appendix

\bibliographystyle{plainurl}

\bibliography{bib}

\section{Hypervolume}

\begin{figure}[ht]
    \centering
    \includegraphics[width=0.2\textwidth]{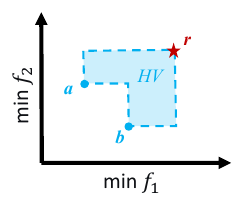} \caption{Illustration of the hypervolume in two‑dimensional minimization case.\label{fig:hv}}
\end{figure}

For a set of solutions $\mathcal{S}$ and a reference point $\bm{r}$, the hypervolume~\cite{zitzler1999multiobjective} $\mathrm{HV}(\mathcal{S},\bm{r})$ measures the portion of the objective space that is dominated by at least one solution in $\mathcal{S}$ and, dominates $\bm{r}$ at the same time.
As shown in Figure~\ref{fig:hv}, in the two‑dimensional minimization case with $f_1$ vs. $f_2$ (where smaller values are better, e.g., MAE vs. multiplicative depth), this corresponds to the total area of the union of the rectangles that have a solution point and $\bm{r}$ as opposite corners.
Given two solutions $\bm{a}$ and $\bm{b}$ (both better than $\bm{r}$ in all objectives), the hypervolume is exactly the area covered by the two rectangles defined by $(\bm{a},\bm{r})$ and $(\bm{b},\bm{r})$, without double‑counting their overlap.
Intuitively, \emph{a larger hypervolume is better}, which indicates that the solution set covers more of the region between the solutions and the reference point, i.e., closer to the true Pareto front.

\section{FHE Inference System Implementation}
We develop our GPU-accelerated FHE inference system\footnote{Code: \url{https://github.com/jianhayes/ATLAS}} based on our extensions of an open-source codebase, i.e., NEXUS-End2End~\cite{nexuse2e}. 
Our experiment platform is a modified version RTX 4090 with 48GB.
The encoder of BERT, decoder of LLaMA, and vision encoder of ViT can be run in constrained 40GB GPU based on our FHE inference system.
As a result, our system's reliance on GPU memory is significantly lower than that of existing frameworks~\cite{zhang2024secure, zhang2025moai, moon2025thor}.
We have also integrated \texttt{LibTorch} into our library, allowing us to build models directly by reading Torch weights, rather than pre-storing large amounts of encoded plaintext weights.
In our system, we can pre-compute the weights packing (this requires ${\sim}200$ GB RAM for BERT), or dynamically pack the weights in runtime.

\subsection{Details of Iterative Softmax in \methodname{}}

\paragraph{Intervals.} Following Cho et al.~\cite{cho2024fast}, the degree of exponential approximation is $15$; the input interval for generating each degree-$(2^{p_j^{(i)}}{-}1)$ polynomial depends on the iteration index $j$: $(0.085, 256)$ for $j{=}1$, $(0.003, 1.238)$ for $j {\in} \{2,3,4\}$, and $(0.0036, 1.0816)$ for $j{=}5$, as shown in Table~\ref{tab:softmax_space}.

\begin{table}[ht]
\centering
\caption{Search space for the iterative softmax approximation. 
The exponential function is approximated by a $15$-degree polynomial. 
The inverse square root in iteration $j$ uses a Chebyshev polynomial of degree $2^{p_j}{-}1$, where $p_1 {\in} [1,7]$ (non‑zero) and $p_2,\dots,p_5 {\in} [0,7]$. 
A zero $p_j$ omits that iteration.}
\label{tab:softmax_space}
\resizebox{0.7\columnwidth}{!}{%
\begin{tabular}{c c c c}
\toprule
 & iter. $j$ & degree & input range \\
\midrule
$\exp(x)$ & all & $15$ & $(-8,\,0)$ \\
\midrule
\multirow{5}{*}{$1/\sqrt{x}$}
& 1 & $2^{p_1}-1$ & $(0.085,\,256)$ \\
& 2 & $2^{p_2}-1$ & $(0.003,\,1.238)$ \\
& 3 & $2^{p_3}-1$ & $(0.003,\,1.238)$\\
& 4 & $2^{p_4}-1$ & $(0.003,\,1.238)$\\
& 5 & $2^{p_5}-1$ & $(0.0036,\,1.0816)$ \\
\bottomrule
\end{tabular}
}
\end{table}

\paragraph{Per-Layer Softmax Max-Constant.} 
%
In FHE, dynamically evaluating $\max_i x_i$ in the softmax exponentiation is prohibitively expensive.
Following prior work~\cite{zhang2024secure}, each per-layer maximum is replaced by a fixed constant in searching and FHE inference, which is calibrated on a single real-data sample using the iterative softmax~\cite{cho2024fast}-based cleartext polynomial model.
We do not jointly search the softmax max‑constant, as it is a data‑dependent property and searching it would overfit to the search data. Instead, we assume that the input already falls within the required range for the iterative softmax with a fixed max-constant.
Empirically, the fixed estimation works well as it matches cleartext accuracy on full validation sets.

For BERT‑Base ($L=12$), one IMDB~\cite{maas2011learning} sentence is used, yielding the 12 integer constants:
\begin{equation}
\bm{c}_{{bert}} = \{\,13, 18, 34, 16, 12, 13, 13, 12, 13, 12, 15, 10\,\},
\end{equation}
for ViT‑Base ($L=12$), one ImageNet‑1K image gives:
\begin{equation}
\bm{c}_{{vit}} = \{\,108, 32, 22, 14, 15, 15, 15, 13, 15, 13, 13, 9\,\},
\end{equation}
for LLaMA3‑8B ($L=32$), the same IMDB procedure yields:
\begin{equation}
\begin{aligned}
\bm{c}_{{llama}} = \{\,&22, 7, 5, 6, 6, 9, 9, 8, 9, 10, 9, 7, 6, 9, 10, 9, 9, 9,\\
                             &13, 11, 12, 10, 8, 13, 9, 9, 10, 9, 12, 11, 11, 89\,\}.
\end{aligned}
\end{equation}
These vectors are injected directly into our softmax computation of all configurations in cleartext simulated evaluation and FHE inference.

\subsection{Details of Polynomials in \methodname{}}

\paragraph{Intervals of Polynomials.}
Fitting polynomials requires a predefined interval. 
Besides the interval for iterative softmax, we also need intervals for normalization and activation functions.
We determine these by using a subset of each task data and the corresponding cleartext model to empirically record the input variance/RMS of normalization layers and the input ranges of GELU/SiLU, then add a 1\%--5\% margin to generate the Chebyshev approximations.
The Chebyshev coefficients of exponential function, invsqrt, GELU, and SiLU used in this work were generated directly by the \texttt{chebyshevform} function in \texttt{Sollya~8.0}, with the intermediate precision set to 1024 bits (for degree 511, to 2048 bits).
After generation, all coefficients with absolute values smaller than $10^{-14}$ were set to zero.
The resulting coefficients were stored in scientific notation, with 40 decimal digits retained for each coefficient.

\paragraph{Polynomial Evaluation with Paterson–Stockmeyer Method.}
Also, we implement the Paterson‑Stockmeyer (PS) polynomial evaluation method in our FHE inference system and apply it uniformly to both power‑basis and Chebyshev‑basis polynomials.
Let $\deg(p)$ be the polynomial degree, for a power‑basis polynomial the multiplicative depth is $\lceil \log_2(\deg(p)+1) \rceil$; for a Chebyshev‑basis polynomial, a single additional linear transformation is required on the input, yielding a depth of $\lceil \log_2(\deg(p)+1) \rceil + 1$.

\subsection{Homomorphic Matrix Multiplication}
For PCMM, we follow the BSGS-based implementation used in NEXUS-EndtoEnd~\cite{nexuse2e}. Since each ciphertext has $32768=128\times256$ slots, we use ct.$(128,128)\times$pt.$(128,128)$ as the basic PCMM operator, which fits within one ciphertext.
%
The ciphertext is first expanded into 16 baby-step rotations. For each of the 16 giant steps, the corresponding plaintext diagonals are encoded and multiplied with the baby-step ciphertexts, and the partial sums are then rotated and accumulated to the final output. Overall, one ct.$(128,128)\times$pt.$(128,128)$ PCMM requires $16\times16=256$ plaintext-ciphertext multiplications and 32 rotations, including 16 baby-step rotations and 16 giant-step rotations. The computation consumes one level.
For higher-dimensional PCMM, we decompose the computation into independent ct.$(128,128)\times$pt.$(128,128)$ block operations and combine the partial results using homomorphic rotations and additions. For ct.$(A,B)\times$pt.$(C,D)$, this gives $(A\cdot B\cdot C)/(2\cdot128^3)$ basic block operations, while the extended PCMM still consumes only one level.
We also extend the codebase~\cite{nexuse2e} to support BSGS-based CCMM using rotations and precomputed masks. 
A ct.$(128,128)\times$ct.$(128,128)$ CCMM consumes three levels due to masking and ciphertext-ciphertext multiplication. 
Unlike PCMM, this CCMM implementation is not dimension-generic, since each new matrix shape requires separately designed masks.

\section{Surrogate Models of LLaMA}\label{sec:llama_surrogate}
\begin{figure}[t]
    \centering
    \begin{subfigure}[b]{0.235\textwidth}
        \centering
        \includegraphics[width=\textwidth]{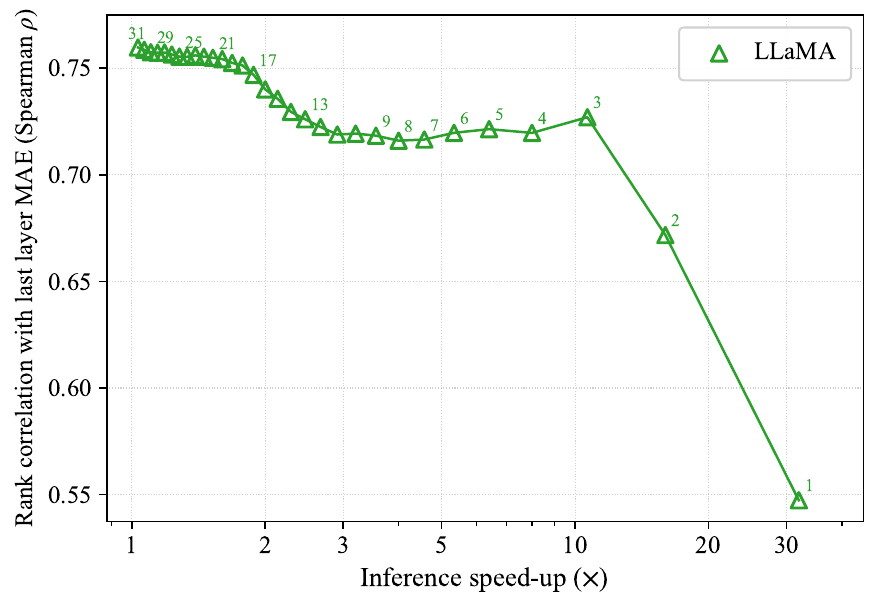}
    \end{subfigure}
    \hfill
    \begin{subfigure}[b]{0.235\textwidth}
        \centering
        \includegraphics[width=\textwidth]{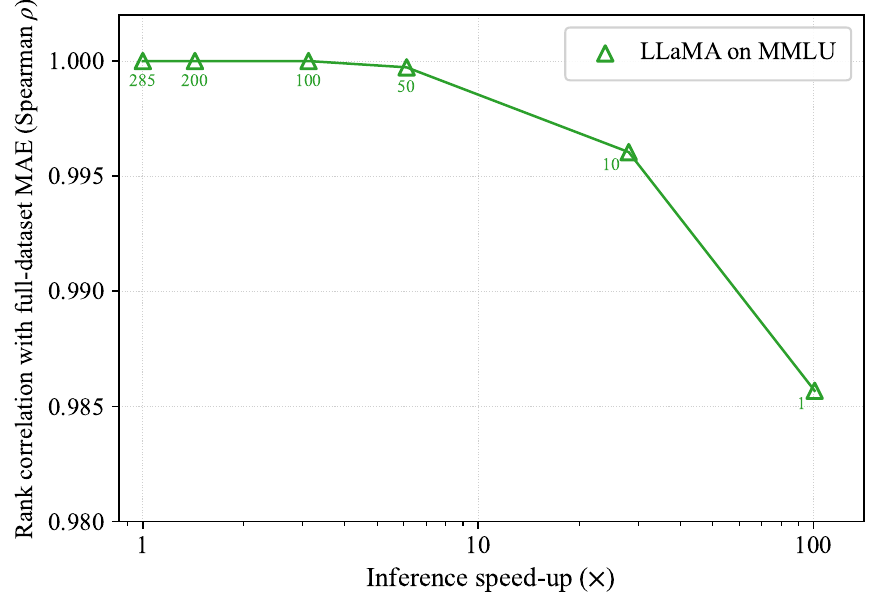}
    \end{subfigure}
    \\
    \begin{subfigure}[b]{0.235\textwidth}
        \centering
        \includegraphics[width=.95\textwidth]{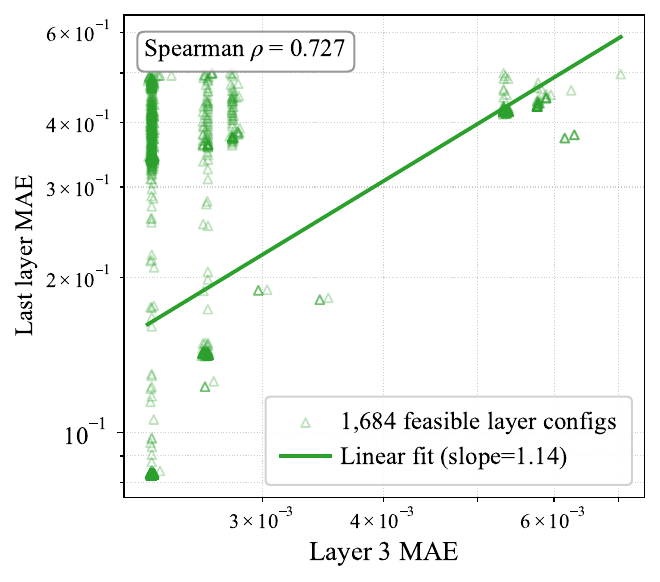}
    \end{subfigure}
    \hfill
    \begin{subfigure}[b]{0.235\textwidth}
        \centering
        \includegraphics[width=.95\textwidth]{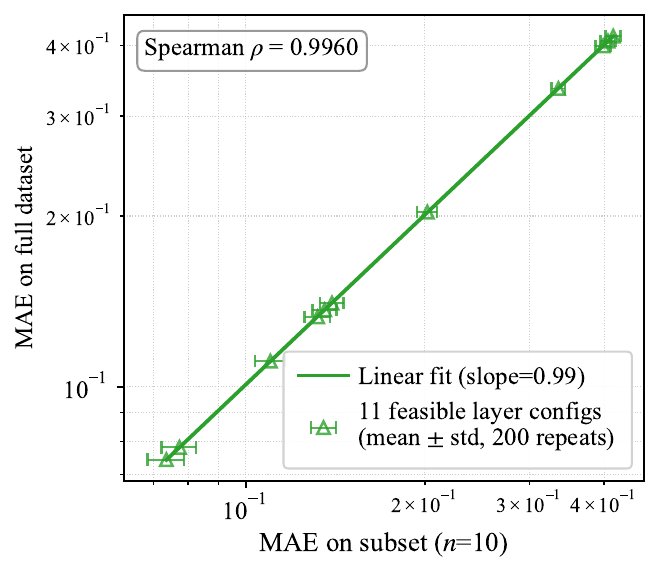}
    \end{subfigure}
    \caption{MAE at an early layer as a surrogate for MAE at the last layer, and the correlation between the MAE on the full and subset of the dataset, are shown on LLaMA.\label{fig:surr_layer_llama}}
\end{figure}

\paragraph{Data Surrogate.} Figure~\ref{fig:surr_layer_llama} (top right) reports Spearman $\rho$ and the resulting speed-up as a function of the proxy subset size for LLaMA3-8B.
As the subset size increases, $\rho$ approaches $1$.
A subset of $10$ MMLU-dev sentences (size of full set is $285$), evaluated on eight \textsc{NetworkProblem} configurations, yields $\rho \approx 0.996$ with a ${\sim}28\times$ speed-up (Figure~\ref{fig:surr_layer_llama}, bottom right), which is sufficient for reliable ranking.
Thus, \methodname{} uses a subset size of $10$ for LLaMA3-8B in practice.

\paragraph{Early-Layer Proxy.} Figure~\ref{fig:surr_layer_llama} (top left) plots $\rho$ between the early layer-$i$ MAE and the last-layer MAE against the corresponding speed-up.
As the layer index increases, $\rho$ rises to $0.75$ but not $1.0$, possibly because the inter‑layer MAE in LLaMA exhibit greater variation, which complicates early-layer proxy.
As illustrated for one example (Figure~\ref{fig:surr_layer_llama}, bottom left), using the layer-3 MAE as the early-layer proxy gives $\rho = 0.727$ with an ${\sim}10.7\times$ speed-up.
Accordingly, \methodname{} adopts the layer-3 MAE as the proxy in LLaMA3-8B.

\section{Extended Experimental Results}\label{sec:appendix_lat}

\paragraph{Accuracy--Latency Pareto Fronts of BERT.}
\begin{figure*}[t]
    \centering
    \begin{subfigure}[b]{0.325\textwidth}
        \centering
        \includegraphics[width=0.985\textwidth]{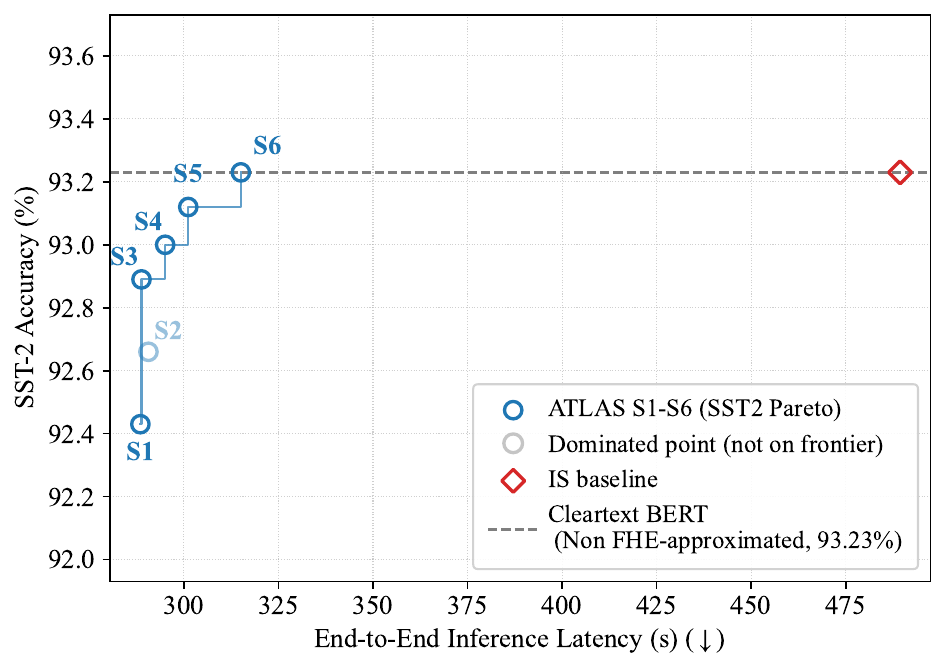}
        \caption{SST-2}\label{fig:fig_bert_lat_sst2}
    \end{subfigure}
    \hfill
    \begin{subfigure}[b]{0.325\textwidth}
        \centering
        \includegraphics[width=0.985\textwidth]{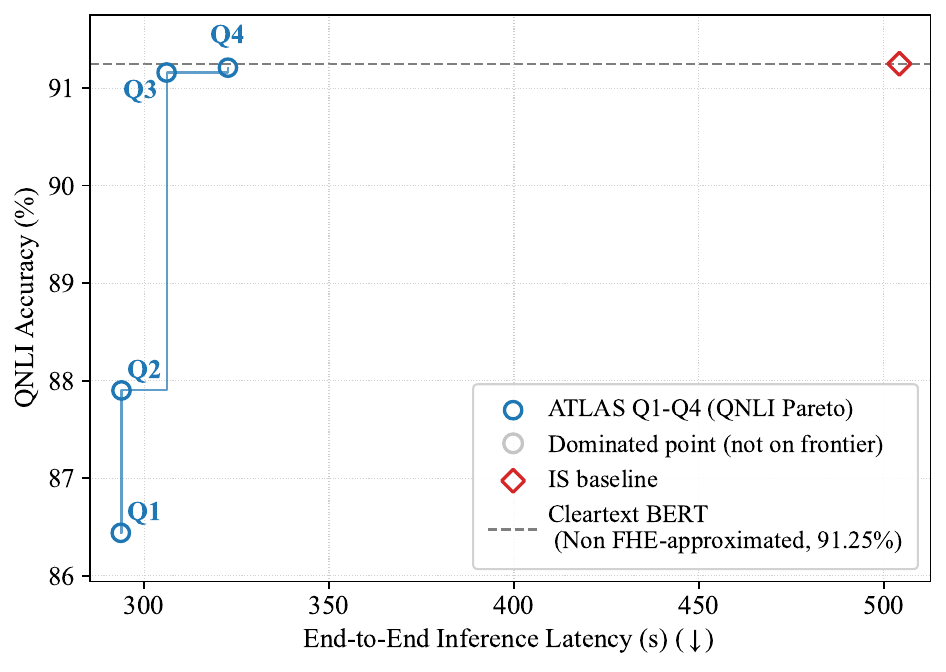}
        \caption{QNLI}\label{fig:fig_bert_lat_qnli}
    \end{subfigure}
    \hfill
    \begin{subfigure}[b]{0.325\textwidth}
        \centering
        \includegraphics[width=0.985\textwidth]{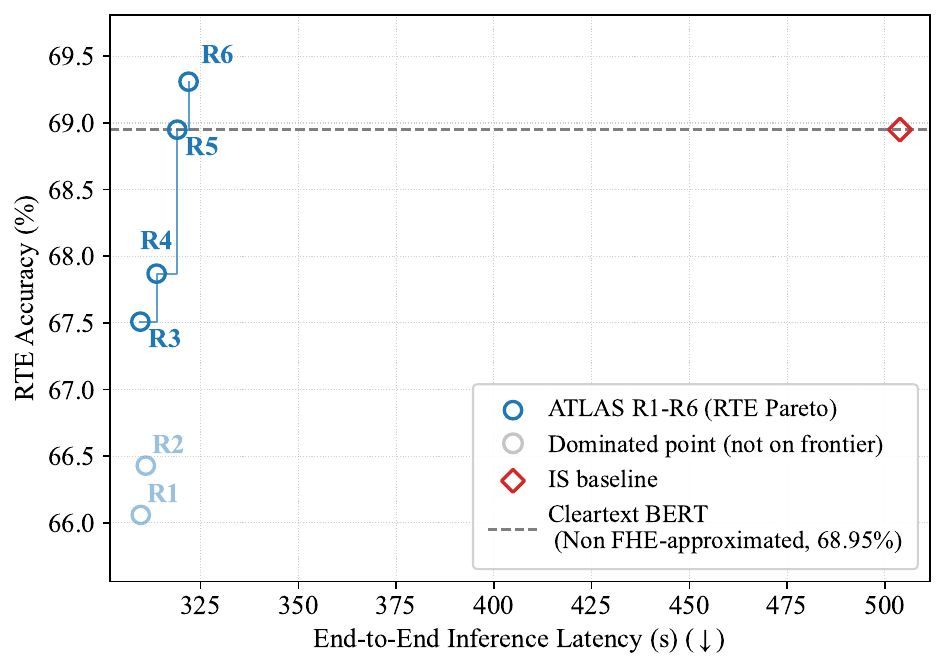}
        \caption{RTE}\label{fig:fig_bert_lat_rte}
    \end{subfigure}
    \caption{Accuracy--latency Pareto fronts discovered by \methodname{} on three tasks of BERT. Each point represents a non-dominated $\bm{\lambda}$ configuration, plotted in the space of end-to-end FHE inference latency vs.\ downstream task accuracy, measured on the \texttt{Desilo-FHE} GPU backend. The iterative softmax baseline is shown as a reference. Dashed horizontal lines indicate the accuracy of the cleartext, non FHE-approximated model. Greyed points are configurations that are non-dominated in depth but dominated once latency is measured.}\label{fig:bert_acc_vs_lat}
\end{figure*}
Figure~\ref{fig:bert_acc_vs_lat} reports the accuracy--latency Pareto fronts discovered by \methodname{} on the three GLUE tasks, measured end-to-end under FHE on the \texttt{Desilo-FHE} GPU backend.
The high-precision iterative softmax baseline (denoted IS) requires $489$~s on SST-2 and $504$~s on QNLI and RTE.
On all three tasks, \methodname{} recovers the accuracy of the underlying finetuned model at roughly two thirds of that cost: S6 attains $93.23\%$ on SST-2 in $315$~s ($35.6\%$ faster), R5 attains $68.95\%$ on RTE in $319$~s ($36.7\%$ faster), and Q4 comes within $0.04$~pp at $91.21\%$ on QNLI in $323$~s ($35.9\%$ faster).
R6 goes further and exceeds the finetuned model at $69.31\%$ on RTE, still $36.1\%$ faster than IS.
The fronts are steep, so accuracy can be traded for additional speed at a favorable rate: on SST-2, S1 gives up $0.80$~pp for a $41.1\%$ reduction ($288$~s), and every configuration from S1 to S6 lands between $288$ and $315$~s against the baseline's $489$~s.
These fronts exhibit the same steep shape observed in the accuracy--depth view of Section~\ref{sec:bert_acc-depth_pareto}.
A configuration that is Pareto-optimal in depth is not automatically Pareto-optimal in latency, however.
The greyed points in Figure~\ref{fig:bert_acc_vs_lat} (S2 on SST-2, R1 and R2 on RTE) are non-dominated in the depth view but dominated once measured end-to-end, because differences of a few levels out of ${\sim}800$ fall within run-to-run latency variation and do not always change the bootstrapping count.

\paragraph{Latency of RTE and QNLI Pareto Fronts.}
Table~\ref{tab:desilo_qnli_rte2} shows the detailed latency of the Pareto configurations Q1--Q4 for QNLI and R1--R6 for RTE on the \texttt{Desilo-FHE} GPU backend. The observed speedups range from ${\sim}36\%$ to ${\sim}42\%$ for QNLI and ${\sim}36\%$ to ${\sim}39\%$ for RTE, respectively. These results provide additional evidence that \methodname{} generalizes effectively across different downstream tasks.
\begin{table}[ht]
\centering
\caption{Latency of QNLI \& RTE BERT Pareto-optimal configurations in \texttt{Desilo-FHE} GPU backend.}
\label{tab:desilo_qnli_rte2}
\resizebox{.7\columnwidth}{!}{%
\begin{tabular}{lcccc}
\toprule
\makecell[l]{Config} & \makecell[c]{Mul.\\Depth} & \#B & \makecell[c]{Lat.\\(s)} & \makecell[c]{Lat.\\Redu.} \\
\midrule
\makecell[l]{IS baseline (QNLI)} & 1212 & 603 & 504 & - \\
\midrule
\methodname{}-Q1 & 756 & 371 & 294 & 41.7\% \\
\methodname{}-Q2 & 757 & 371 & 294 & 41.7\% \\
\methodname{}-Q3 & 801 & 387 & 306 & 39.3\% \\
\methodname{}-Q4 & 827 & 399 & 323 & 35.9\% \\
\midrule
\makecell[l]{IS baseline (RTE)} & 1212 & 603 & 504 & - \\
\midrule
\methodname{}-R1 & 816 & 416 & 310 & 38.5\% \\
\methodname{}-R2 & 818 & 416 & 311 & 38.3\% \\
\methodname{}-R3 & 820 & 420 & 310 & 38.5\% \\
\methodname{}-R4 & 839 & 420 & 314 & 37.7\% \\
\methodname{}-R5 & 848 & 424 & 319 & 36.7\% \\
\methodname{}-R6 & 877 & 436 & 322 & 36.1\% \\
\bottomrule
\end{tabular}%
}
\end{table}

\paragraph{Latency of ViT Pareto Fronts.}
In Table~\ref{tab:pareto_lat_vit}, we also implement a ViT demonstration in our inference system. The latency evaluation of the ViT configurations V1--V3 yields a speedup of ${\sim}14\%$, which is smaller than the speedup observed for BERT. This difference is expected, as {optimal ViT requires a greater depth than BERT}, leaving less room for approximation-induced latency reduction. 
Nevertheless, these results still confirm that \methodname{} can deliver meaningful acceleration across different model architectures.
\begin{table}[ht]
\centering
\caption{Latency of ViT Pareto-optimal configurations in \methodname{} inference system based on \texttt{Phantom-FHE} backend.}
\label{tab:pareto_lat_vit}
\resizebox{0.65\columnwidth}{!}{%
\begin{tabular}{lcccc}
\toprule
\makecell[l]{Config} & \makecell[c]{Mul.\\Depth} & \#B & \makecell[c]{Lat.\\(s)} & \makecell[c]{Lat.\\Redu.} 
\\
\midrule
IS baseline & 1212 & 1728 & 2654 & - \\
\midrule
\methodname{}-V1 & 1067 & 1392 & 2288 & 13.8\% \\
\methodname{}-V2 & 1071 & 1374 & 2267 & 14.6\% \\
\methodname{}-V3 & 1073 & 1362 & 2259 & 14.9\% \\
\bottomrule
\end{tabular}%
}
\end{table}

\begin{figure}[ht]
    \centering
    \begin{subfigure}[b]{0.235\textwidth}
        \centering
        \includegraphics[width=\textwidth]{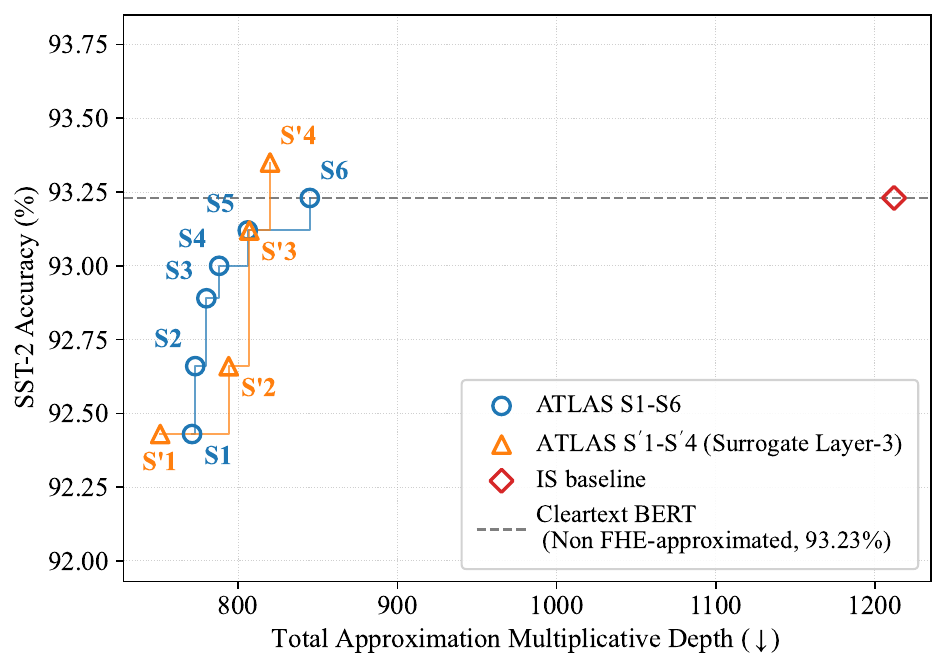}
        \caption{Accuracy--depth Pareto}\label{fig:fig_earlylayer_acc_vs_depth}
    \end{subfigure}
    \hfill
    \begin{subfigure}[b]{0.235\textwidth}
        \centering
        \includegraphics[width=\textwidth]{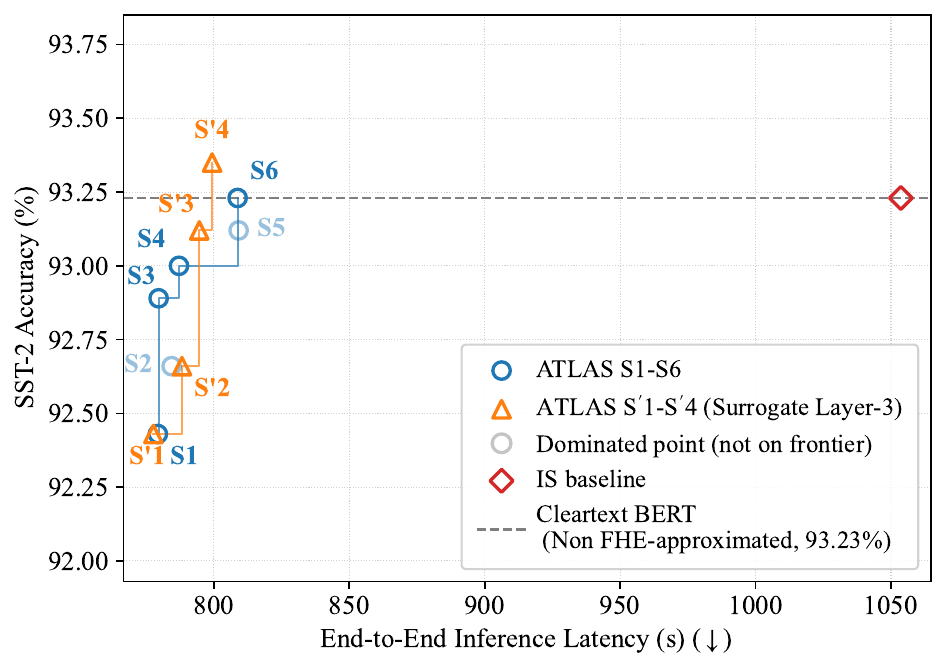}
        \caption{Accuracy--latency Pareto}\label{fig:fig_earlylayer_acc_vs_lat}
    \end{subfigure}
    \caption{
    Early-layer proxy comparison on BERT/SST-2.
    }\label{fig:earlylayer_comparison}
\end{figure}

\paragraph{Early-layer Proxy Pareto Front.}
Early-layer proxies can discover Pareto fronts that closely match those obtained with the full model.
For example, Figure~\ref{fig:earlylayer_comparison} shows that for BERT/SST-2, the layer-3 proxy produces S'1--S'4 with depth and end-to-end latency (in \texttt{Phantom-FHE}) close to S1--S6, confirming its effectiveness in speeding up the search.

\end{document}